\documentclass[twocolumn,superscriptaddress,pra,longbibliography]{revtex4-2}
\usepackage{amsmath,amsfonts,amssymb}
\usepackage{graphicx}
\usepackage{xcolor}
\usepackage{xr-hyper}
\usepackage[colorlinks=true,urlcolor=blue,citecolor=blue,linkcolor=blue]{hyperref}

\begin{document}
\newcommand{\iac}[1]{{\color{blue}#1}}

\newcommand{\tom}[1]{{\color{red}#1}}
\newcommand{\iaciac}[1]{{\color{magenta}#1}}

\title{Mean-chiral displacement in coherently driven photonic lattices and its application to synthetic frequency dimensions}

\author{Greta Villa}
\affiliation{Department of Physics, University of Konstanz, 78464 Konstanz, Germany}
\affiliation{Pitaevskii BEC Center, CNR-INO and Dipartimento di Fisica, Universit\`a di Trento, I-38123 Trento, Italy}
\affiliation{Advanced Institute for Materials Research (WPI-AIMR), Tohoku University, Sendai 980-8577, Japan}

\author{Iacopo Carusotto}
\affiliation{Pitaevskii BEC Center, CNR-INO and Dipartimento di Fisica, Universit\`a di Trento, I-38123 Trento, Italy}

\author{Tomoki Ozawa}
\email{tomoki.ozawa.d8@tohoku.ac.jp}
\affiliation{Advanced Institute for Materials Research (WPI-AIMR), Tohoku University, Sendai 980-8577, Japan}

\date{\today}
\begin{abstract}
    Characterizing topologically nontrivial photonic lattices by measuring their topological invariants is crucial in topological photonics. In conservative one-dimensional systems,  a widely used observable to extract the winding number is the mean-chiral displacement. In many realistic photonic systems, however, losses can hardly be avoided, and little is known on how one can extend the mean-chiral displacement to a driven-dissipative context. Here we theoretically propose an experimentally viable method to directly detect the topological winding number of one-dimensional chiral photonic lattices. The method we propose is a generalization of the mean-chiral displacement to a driven-dissipative context with coherent illumination. By integrating the mean-chiral displacement of the steady state over the pump light frequency, one can obtain the winding number with a correction of the order of the loss rate squared. We demonstrate that this method can be successfully applied to lattices along synthetic frequency dimensions.
\end{abstract}
\maketitle

\section{Introduction}
Topological photonics, based on the exploration of the topological structure of photonic states in suitable parameter spaces, has been successful in providing powerful methods to engineer exotic photonic band structures and robust boundary modes~\cite{Lu:2014:NatPhot,Ozawa:2019:RMP,Ota:2020:Nanophot,Segev:2020:Nanophot,Chen:2021:Elight,Price:2022:Roadmap,Zhang:2023:Nature}. Arguably, the simplest topological structures one can construct consist of one-dimensional lattices with a chiral symmetry~\cite{Asboth:2016:Lecture}. The most celebrated example in this class is the Su-Schrieffer-Heeger (SSH) model, which was originally proposed as a model to describe the electronic properties of polyacetylene~\cite{Su:1979:PRL}. The bulk eigenmodes of such chiral lattices are characterized by an integer-valued topological invariant called the winding number $\mathcal{W}$. Correspondingly, there exist $\mathcal{W}$ edge localized modes under the open boundary condition, which is a manifestation of the more general principle called the bulk-edge correspondence~\cite{Chiu:2016:RMP}.

The experimental confirmation of a non-trivial topology is often obtained through the detection of the edge-localized modes~\cite{Ozawa:2019:RMP}. From this, through the bulk-edge correspondence, one infers that the bulk eigenmodes are topological. In some cases, rather than relying on the bulk-edge correspondence, it is useful to characterize the topological features of the model by directly detecting the winding number, e.g. to explicitly confirm the bulk-edge correspondence. In the last decade, strategies of this kind have been investigated both theoretically~\cite{Ozawa:2014:PRL,bardyn2014measuring} and experimentally~\cite{Wimmer:2017:NatPhys,Gianfrate:2020:Nature} for different two-dimensional topological models. Such a development is specially relevant in systems that do not display a well-defined boundary, in which case the detection of the edge-localized modes is not possible and one has no other option but to directly look at the bulk topology.

Measurement of the topological winding number $\mathcal{W}$ of one-dimensional chiral lattices from the bulk eigenstate wavefunction typically requires extracting the relative phase of the wavefunction between the two sublattices, which is often experimentally challenging. A powerful method to obtain $\mathcal{W}$ from the intensity profiles only, is to measure the mean-chiral displacement~\cite{Cardano:2017:NatComm,Maffei:2018:NJP,Longhi:2018:OptLett}: this is the expectation value of the operator $\Gamma x$, where $\Gamma$ is the chiral operator and $x$ is the position operator. For an initial state localized on the central unit cell at $x = 0$, the mean-chiral displacement $\langle \Gamma x\rangle$ converges to $\mathcal{W}/2$ in the long-time limit of a conservative evolution. Measurement of the mean chiral displacement has been successfully implemented to detect the winding number in several photonic platforms~\cite{Cardano:2017:NatComm,Jiao:2021:PRL,caceres2023edge}.

In all these works, application of the original formulation of the mean chiral displacement method was possible thanks to the reduced amount of losses, which made the propagation of the light field to be accurately described in terms of a unitary evolution.
Since losses are an unavoidable feature of various photonic setups, it is desirable to find alternative methods to measure the winding number in more general cases where losses are significant. A pioneering step in this direction was made by the experiment~\cite{StJean:2021:PRL} where the winding number of polaritonic lattices was measured under a continuous-wave incoherent illumination.

Here we make a further step by theoretically considering systems under a coherent monochromatic illumination. 
In specific, we show that a frequency-integration of the mean-chiral displacement measured on the steady-state at a given pump frequency provides an accurate estimate of the winding number in realistic cases where losses are comparable or smaller than the characteristic bandwidth of the photonic states. Our method based on the frequency-integrated mean chiral displacement keeps being accurate for relatively small lattice sizes, providing a versatile tool to measure the topological winding number in generic driven-dissipative photonic systems.

As a specific example of application, we discuss how our method can be successfully applied to the emerging platform of synthetic frequency dimensions~\cite{Ozawa:2016:PRA,Yuan:2016:OptLett,Yjan:2018:Optica,Ozawa:2019:NatRevPhys,Dutt:2019:NatComm,Lustig:2019:Nature,Dutt:2020:Light,Dutt:2020:Science,Wang:2021:Science,Yuan:2021:APL,Lustig:2021:Adv,Dutt:2022:NatComm,Balcytis:2022:SciAdv,Ehrhardt:2023:LPR} where it may provide an efficient way of probing the lattice topology. This is all the way more relevant as the incoherent pump scheme of Ref.~\cite{StJean:2021:PRL} is expected to suffer from an intrinsic difficulty related to the broadband nature of the drive which is hardly able to selectively address a single lattice cell as required in the mean chiral displacement approach.
As compared to recent experiments in this context that are based on a wavefunction tomography method~\cite{Li:2023:Light,Pellerin:PRL2024}, our proposed method directly addresses the physical consequences of the winding number rather than extracting it from the tomography of bulk band states. As such, its implementation would provide a complementary information on the bulk topology of the photonic lattice.

\section{Results}
\subsection{One dimensional chiral Hamiltonian and the Mean-chiral displacement}
In this first Section, we briefly review the basic concepts of chiral one-dimensional lattices~\cite{Asboth:2016:Lecture} and mean-chiral displacement~\cite{Cardano:2017:NatComm,Maffei:2018:NJP} and we introduce the basic terminology that will be used in the rest of the paper.

\subsubsection{One dimensional chiral Hamiltonian}
We consider one-dimensional lattice models with chiral symmetry, that is, the entire lattice can be divided into A and B sublattices, and particles (photons) can hop between sites in different sublattices but not within the same sublattice. We focus on the case where a unit cell consists of two sites, one from each sublattice, which is relevant for the SSH model. 
We denote by $|x,A\rangle$ ($|x,B\rangle$) a state where a particle is in sublattice A (B) in $x$-th unit cell,
the integer $x$ ranging between $-\infty$ and $+\infty$ for an infinite lattice.

In the general case, the real-space Hamiltonian of one-dimensional tight-binding models with chiral and translational symmetries takes the following form
\begin{align}
    H = \sum_{x,j} J_j |x+j,A\rangle \langle x,B| + \mathrm{H.c.},
\end{align}
where $j$ runs through all, positive and negative integer numbers and the hopping amplitudes $J_j$ are taken to be real numbers (extension to complex $J_j$ would be straightforward).
Because of the translational symmetry (i.e. $J_j$ does not depend on $x$), the Hamiltonian is diagonal in momentum space labeled by momentum $k$.
In the momentum-space basis, we have
\begin{align}
    |x,A\text{ (or B)}\rangle = \frac{1}{\sqrt{2\pi}}\int_0^{2\pi}dk\, e^{-ikx}|k,A\text{ (or B)}\rangle,
\end{align}
and the Hamiltonian takes the form
\begin{align}
    H = \int_0^{2\pi}dk
    \begin{pmatrix}
    | k,A \rangle & | k,B\rangle
    \end{pmatrix}
    H(k)
    \begin{pmatrix}
    \langle k,A | \\ \langle k,B |
    \end{pmatrix},
\end{align}
in terms of the momentum-space Hamiltonian
\begin{align}
    H(k) =
    \begin{pmatrix}
    0 & \sum_j e^{-ijk} J_j \\ 
    \sum_j e^{ijk} J_j & 0
    \end{pmatrix}. \label{eq:hamk}
\end{align}
Splitting the off-diagonal element in its modulus and the phase
\begin{align}
    \sum_j e^{ijk} J_j \equiv E(k) \,e^{i\theta(k)},
\end{align}
where $E(k) \ge 0$ and $\theta (k)$ are real functions with periodicity of $2\pi$ in $k$,
the eigenvalues and eigenvectors of $H(k)$ are respectively $E_\pm(k)=\pm E(k)$ and
\begin{align}
    |u_\pm (k) \rangle =
    \frac{1}{\sqrt{2}}
    \begin{pmatrix}
        e^{-i\theta (k)} \\ \pm 1
    \end{pmatrix} \label{eq:uk}
\end{align}
and physically correspond to the Bloch energy and the Bloch wavefunction in the bulk of the photonic lattice.
In momentum space, the chiral symmetry of the lattice translates into the momentum-space Hamiltonian obeying $\Gamma H(k) \Gamma^\dagger = -H(k)$ with
\begin{align}
    \Gamma \equiv
    \begin{pmatrix}
    1 & 0 \\ 0 & -1    
    \end{pmatrix}.
\end{align}
As a result, the eigenstates are related by this chiral symmetry, $|u_\mp (k) \rangle = \Gamma |u_\pm (k)\rangle$ and their eigenenergies $\pm E(k)$ are symmetric around 0.

For these chiral one-dimensional lattices, the topological invariant of $H(k)$ is given by the integer winding number, defined by
\begin{align}
    \mathcal{W} = \frac{1}{2\pi}\int_0^{2\pi} dk \,\frac{d \theta(k)}{d k}:
\end{align}
this can be understood as the number of times the phase of the off-diagonal element of $H(k)$ changes by $2\pi$ as one varies $k$ from 0 to $2\pi$, which equals to the number of times the off-diagonal element of $H(k)$ winds around the origin in the complex plane. Next we will see the winding of the off-diagonal element for some concrete models. 

\subsubsection{Specific examples with $\mathcal{W} = 0, 1, 2$}
\label{sec:examples}

To make our discussion more concrete, we now introduce three specific examples of one-dimensional chiral lattice Hamiltonians featuring winding numbers $\mathcal{W}=0$, $1$, and $2$. These examples will be used  in the rest of the paper to illustrate the efficiency of our proposal on specific cases.

\begin{figure*}[htbp]
    \centering
    \includegraphics[width=0.8\textwidth]{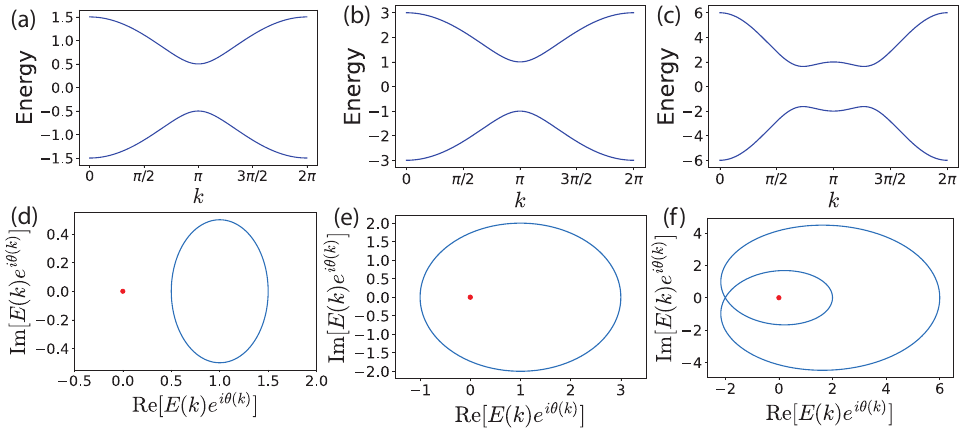}
    \caption{Band structures and off-diagonal component of $H(k)$ for three different models of one-dimensional chiral Hamiltonians. Band structures are shown in panels (a, b, c), and the off-diagonal component of $H(k)$ are shown in panels (d,e,f). Vertical axis of (a,b,c) and both axes of (d,e,f) are in units of $J_0$. Panels (a) and (d) are for the model with $J_1 = 0.5 J_0$ and $J_2 = 0$, which has a winding number $\mathcal{W}=0$. Panels (b) and (e) are for the model with $J_1 = 2 J_0$ and $J_2 = 0$, which has a winding number $\mathcal{W}=1$. Panels (c) and (f) are for the model with $J_1 = 2J_0$ and $J_2 = 3 J_0$, which has a winding number $\mathcal{W}=2$.
    For each model, the winding number $\mathcal{W}$ can be read off from panels (d,e,f) as the number of times the off-diagonal element of $H(k)$ winds around the origin of the complex plane (indicated by red dots).}
    \label{fig:examples}
\end{figure*}

Lattices with winding numbers $\mathcal{W}=0$ and $1$ can be obtained by just considering nearest-neighbor hoppings, i.e. by setting $J_{|j|\geq 2}=J_{-1}=0$ and considering only $J_{0,1}$ to be non-zero, which is often called the Su-Schrieffer-Heeger (SSH) model.
As it is known from the theory of the SSH model~\cite{Asboth:2016:Lecture}, the winding number is $\mathcal{W}=0$ when $J_0 > J_1$ and $1$ when $J_0 < J_1$. To obtain larger values, we need to include longer-range hoppings: for example, a $\mathcal{W}=2$ can be constructed by supplementing the non-vanishing $J_{0,1}$ with  a non-zero  and large enough $J_2$.

In Fig.~\ref{fig:examples}, we plot in panels (a,b,c) the band structure of three specific models yielding $\mathcal{W}=0$, $1$, and $2$ and, in panels (d,e,f), we show the corresponding plots of  the off-diagonal element of $H(k)$ in the complex plane. In these latter plots, the winding around the origin of the complex plane as one sweeps the momentum $k$ through the Brillouin zone gives the winding number $\mathcal{W}$ of the photonic lattice.
Fig.~\ref{fig:examples}(a,d) are for the model with $J_1 = 0.5 J_0$ and $J_2 = 0$, for which $\mathcal{W}=0$.
Fig.~\ref{fig:examples}(b,e) are for the model with $J_1 = 2 J_0$ and $J_2 = 0$,  for which $\mathcal{W}=1$.
Fig.~\ref{fig:examples}(c,f) are for the model with $J_1 = 2 J_0$ and $J_2 = 3J_0$,  for which $\mathcal{W}=2$.
These specific examples will be used for our discussion in the later Sections of the work.

\subsubsection{Mean-chiral displacement}
In conservative systems, a powerful method to experimentally access the winding number is through the measurement of the mean chiral displacement. One first starts from an initial condition localized in the central unit cell, and then lets the system evolve in time according to the Hamiltonian. The expectation value of the mean chiral operator $\langle \Gamma x\rangle$ can be shown to approach $\mathcal{W}/2$ in the long time limit. 

To briefly prove this statement, we  write the state vector in the position basis as
\begin{align}
    |\psi (t)\rangle = \sum_x \left\{ \psi(x,A;t)|x,A\rangle + \psi(x,B;t)|x,B\rangle \right\},
\end{align}
where $x$ is an integer quantity ranging between $\pm\infty$. We also define the state vector at $x$-th unit cell as a two-component spinor
\begin{align}
    |\psi (x;t)\rangle =
    \begin{pmatrix}
    \psi (x,A;t) \\
    \psi (x,B;t)
    \end{pmatrix}.
\end{align}
At $t = 0$, we assume that the wavefunction is completely localized in the central unit cell at $x = 0$, namely the wavefunction is
\begin{align}
    |\psi (x;0)\rangle =
    \delta_{x,0}
    |u_0 \rangle,
\end{align}
where $|u_0\rangle$ is the (arbitrarily chosen) initial state profile in the unit cell at $x=0$ satisfying the normalization $\langle u_0 | u_0\rangle = 1$.
The wavefunction at a later time $t$ is then given by $|\psi (t)\rangle = e^{-iHt}|\psi (0)\rangle$.

Noting that the Bloch eigenstates $e^{ikx}|u_\pm (k)\rangle$ form a complete set of states, we can expand $|\psi (x;0)\rangle = \delta_{x,0}|u_0\rangle$ in terms of these basis states. The wavefunction at a generic later time $t$ can then be written as
\begin{align}
    |\psi (x;t)\rangle
    &= \frac{1}{2\pi}\sum_{\alpha=\pm}\int_0^{2\pi}dk e^{-i E_\alpha (k) t} e^{ikx}|u_{\alpha} (k)\rangle \langle u_\alpha (k)|u_0\rangle.
\end{align}

From this expression, the average value of the mean chiral operator can be then calculated as
\begin{align}
    &\langle \Gamma x \rangle (t)
    =
    \sum_x \langle \psi (x;t) | \Gamma x |\psi (x;t)\rangle
    \notag \\
    &=
    \frac{1}{2\pi}\int_0^{2\pi} dk \left[ \frac{1}{2}\frac{d\theta(k)}{dk} + (\text{oscillating terms}) \right],
\end{align}
where the ``oscillating terms'' involve the product of an oscillating factor $e^{\pm 2iE(k)t}$ times a quantity that is independent of $t$ and periodic in $k \to k + 2\pi$. The integral over $k$ of such oscillating terms tends to vanish in the long-time limit $t\to\infty$ as the frequency of the oscillations  $e^{\pm 2iE(k)t}$ gets faster for growing $t$. 

One therefore obtains the final expression
\begin{align}
    \lim_{t \to \infty} \langle \Gamma x \rangle (t) = \mathcal{W}/2.
\end{align}
relating the mean chiral displacement to the winding number in the photonic lattice. This result holds independently of the specific form $|u_0\rangle$ of the initial state, provided it is localized within the central unit cell at $x=0$.

From the experimental point of view, the significance of this formula stems from the fact that the measurement of the mean chiral displacement does not involve the phase of the wavefunction, and it can thus be extracted from the intensity distribution only. As such it is straightforwardly accessible in most systems without having to rely on  interference effects. While the theory reviewed in this Section refers to conservative systems, in the next Section we are going to generalize the mean chiral displacement method to driven-dissipative systems, with a special eye to coherently driven ones.

\subsection{Mean-chiral displacement under a coherent illumination}

As we have reviewed in the previous Section, the mean-chiral displacement provides an invaluable way to measure the winding number in conservative systems. In many cases relevant to topological photonics, however, the dynamics is not a conservative one but suffers from significant photon losses stemming from radiative and/or non-radiative absorption processes. In this case, one does not have experimental access to the late-time part of the time-dependent dynamics when the light intensity has dropped to very small values. One is rather interested in looking at the steady state reached by the system as a result of the interplay of a continuous driving and of losses. A recent work~\cite{StJean:2021:PRL} has shown that the mean chiral displacement provides an efficient probe in those incoherent pumping schemes that can be implemented, e.g., in polaritonic lattices. Here we complete the picture by demonstrating that a similar scheme also works when coherent pump illuminations are considered, as it is typically the case of experiments using dielectric-based photonic lattices or synthetic frequency dimensions. 

The basic idea behind our proposal is the following. The mean-chiral displacement method can be regarded as being based on the time evolution  in response to a pulsed source $\propto \delta_{x,0}\delta(t)$. Since $\delta (t) = \int_{-\infty}^\infty d\omega e^{-i\omega t}/(2\pi)$ contains all frequencies, we can alternatively consider this as a superposition of the time evolution under coherent drives of different frequencies $\omega$. On this basis, we expect that the topological winding number may be extracted as an integral over the coherent drive frequency: for each value of the drive frequency, the system quickly converges to a time-independent steady-state and it is thus straightforward to measure the mean chiral displacement from the intensity distribution. In the following of this Section, we show how this naive working hypothesis provides indeed quantitatively correct results in the regime of small losses and large systems.

In the presence of a coherent pump and of uniform loss, the equation of motion describing the system is
\begin{align}
    i\frac{\partial}{\partial t}|\psi (t)\rangle = \left( H -i\gamma \right) |\psi(t)\rangle + |s(t)\rangle. \label{eq:schdd}
\end{align}
In particular, we consider a configuration where the source oscillates coherently at a frequency $\omega$ as $|s(t)\rangle = |s\rangle e^{-i\omega t}$ and we look for the steady state which oscillates at the same frequency $|\psi(t)\rangle = |\psi_\omega \rangle e^{-i\omega t}$. The time-independent steady state $|\psi_\omega\rangle$ can be obtained by solving the following matrix equation:
\begin{align}
    |\psi_\omega \rangle = \left( \omega + i\gamma -  H \right)^{-1} |s \rangle. \label{eq:steadygeneral}
\end{align}

We further assume that the source is localized within the central  $x=0$ unit cell, $|s(x;t)\rangle = |s\rangle_0 \delta_{x,0} e^{-i\omega t}$, and we look for the steady state $|\psi (x;t)\rangle = |\psi_\omega (x) \rangle e^{-i\omega t}$. Expanding Eq.~(\ref{eq:steadygeneral}) over the complete basis of Bloch states $e^{ikx}|u_\alpha (k)\rangle$, one obtains
\begin{align}
    |\psi_\omega (x)\rangle = \frac{1}{2\pi}\sum_{\alpha = \pm}\int dk \frac{\langle u_\alpha (k)|s\rangle_0}{\omega + i\gamma - E_\alpha(k)} e^{ikx}|u_\alpha (k)\rangle. \label{eq:steady}
\end{align}
In order to extract the winding number, the mean chiral displacement evaluated on this steady state, $\sum_x \langle \psi_\omega (x)| \Gamma x | \psi_\omega (x)\rangle$, should be integrated over the frequency $\omega$, 
\begin{align}\label{eq:Gammax}
    \langle \Gamma x\rangle_\mathrm{int}
    =
    \frac{\int_{-\infty}^\infty d\omega \sum_x \langle \psi_\omega (x) | \Gamma x | \psi_\omega (x)\rangle}
    {\int_{-\infty}^\infty d\omega \sum_x \langle \psi_\omega (x) | \psi_\omega (x)\rangle},
\end{align}
where the denominator is included for a proper normalization.

After some algebra,
one can show that the integrated mean-chiral displacement is equal to
\begin{align}
    \langle \Gamma x\rangle_\mathrm{int}
    =
    \frac{1}{4\pi}\int dk \frac{d\theta(k)}{dk}\frac{E(k)^2}{E(k)^2 + \gamma^2}.
\end{align}
When the loss $\gamma$ is smaller than the typical energy scale for the hopping amplitude, this formula can be approximated as 
\begin{align}
    \langle \Gamma x\rangle_\mathrm{int}
    &\approx
    \frac{1}{4\pi}\int dk \frac{d\theta(k)}{dk}\left( 1 - \frac{\gamma^2}{E(k)^2}\right) \notag \\
    &=
    \frac{\mathcal{W}}{2} - \frac{\gamma^2}{4\pi}\int dk \frac{d\theta(k)}{dk}\frac{1}{E(k)^2}. \label{eq:mainresult}
\end{align}
This is the central result of our work: as in the long-time limit of the mean-chiral displacement in conservative systems, the leading term  exactly recovers the winding number $\mathcal{W}/2$. Furthermore, as in the incoherently pumped system of Ref.~\cite{StJean:2021:PRL}, the next correction in the small parameter $\gamma$ is roughly proportional to $\gamma^2$ over the squared photonic lattice bandwidth.
We note that this result is independent of the choice of the source profile $|s\rangle_0$ within the central unit cell.

\begin{figure}
    \centering
    \includegraphics[width=0.5\textwidth]{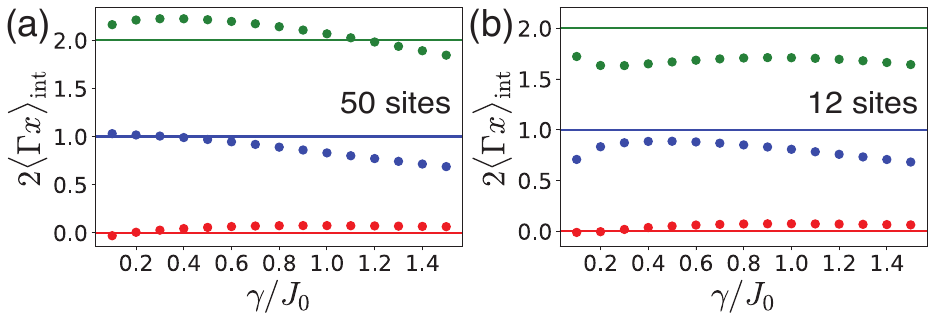}
    \caption{Numerical simulation of $2 \langle \Gamma x\rangle_\mathrm{int}$ as a function of the loss $\gamma$ for two different lattice sizes. (a) 50 lattice sites (25 unit cells) and (b) 12 lattice sites (6 unit cells). For both panels, we numerically integrated the mean chiral displacement over the pump frequency $\omega$ from $-5J_0$ to $+5J_0$ with increments of $\Delta \omega = 0.5J_0$. The bottom red dots, the middle blue dots, and the top green dots refer to the three models illustrated in panels (1a,1d), (1b,1e), and (1c,1f), respectively, in Fig.\ref{fig:examples}.
    The horizontal lines are guides to the eye indicating the actual winding number of the photonic lattice.}
    \label{fig:2gx}
\end{figure}

To put our result on more concrete and quantitative grounds, we numerically evaluate $2 \langle \Gamma x \rangle_\mathrm{int}$ for the specific models with different $\mathcal{W}=0$, 1, and 2 illustrated in Fig.\ref{fig:examples}. A source profile $|s\rangle_0 = (1,0)^T$ was considered, but, provided the source is restricted to the central unit cell and the lattice is long enough for boundary effects to be negligible, the results are independent of this specific choice. The results of the calculations are shown in Fig.~\ref{fig:2gx}.

In panel (a), we plot the numerical prediction for $2 \langle \Gamma x \rangle_\mathrm{int}$ as a function of the loss rate $\gamma$ in a lattice of 50 sites (25 unit cells) with open boundary conditions. The bottom, middle, and top dots are the results for the models with winding number $\mathcal{W}=0$, 1, and 2, respectively. For such a large lattice, we find good agreement for a relatively wide range of $\gamma$. The deviation starts being noticeable when $\gamma$ gets comparable or larger than the characteristic hopping amplitudes $J_j$.

In view of experiments, our method keeps being efficient also for relatively small lattice sizes. This is illustrated in panel (b), where we plot the result of simulations of $2 \langle \Gamma x \rangle_\mathrm{int}$ for a much shorter chain of 12 lattice sites (6 unit cells). Compared to the simulation in panel (a), the agreement between $2 \langle \Gamma x \rangle_\mathrm{int}$ and the winding number $\mathcal{W}$ is certainly worse due to finite size effects, but one can still obtain a reasonable estimate of $\mathcal{W}$. This shows the robustness of the method. 

For the sake of completeness, we note that two separate and somehow competing factors need to be considered to understand the origin of the deviation of the integrated mean chiral displacement from the actual winding number. One such effect is encoded in the second term of Eq.~(\ref{eq:mainresult}): this correction monotonically grows with $\gamma$. The other one is a finite-size effect and is related to the fact that the pumped light can travel to the edges of the system and then be reflected back towards inside: as the propagation length in the lattice decreases with $\gamma$, this latter effect is more pronounced for small $\gamma$ and, of course, for small lattice sizes. Its presence explains why the actual winding number is not recovered in the small $\gamma$ limit in the figure; this deviation is of course weaker for larger lattices and disappears in the infinite-size limit. 

\subsection{Mean chiral displacement along a synthetic frequency dimension}
A platform attracting a growing interest in the photonic community for topological band engineering is based on the so-called synthetic frequency dimensions scheme~\cite{Ozawa:2016:PRA,Yuan:2016:OptLett,Yjan:2018:Optica,Ozawa:2019:NatRevPhys,Dutt:2019:NatComm,Lustig:2019:Nature,Dutt:2020:Light,Dutt:2020:Science,Wang:2021:Science,Yuan:2021:APL,Lustig:2021:Adv,Dutt:2022:NatComm,Balcytis:2022:SciAdv,Ehrhardt:2023:LPR}. The key idea of this approach is to use the different modes of an optical cavity as an extra dimension, so that the different sites along the synthetic dimensions can be selectively addressed via their frequency.  In this Section, we first briefly review~\cite{Dutt:2020:Light,Li:2023:Light,Pellerin:PRL2024} the operation principle of a SSH lattice in a frequency synthetic dimension platform and, then, we elucidate how one can extract the topological winding number from a measurement of the mean chiral displacement along the frequency direction.

\begin{figure}
    \centering
    \includegraphics[width=0.45\textwidth]{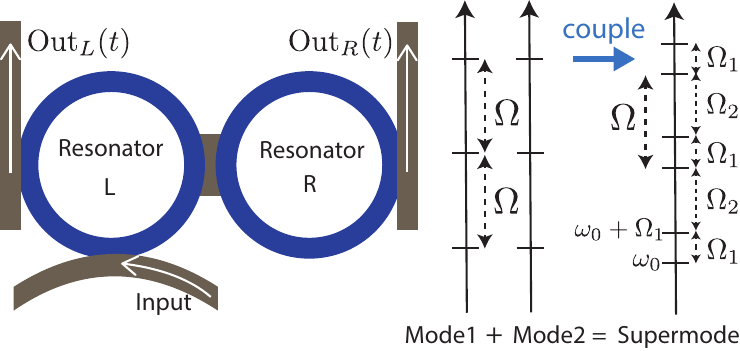}
    \caption{Schematic illustration of two coupled resonators. This setup gives rise to a mode profile which can be used to realize the SSH model in the frequency synthetic dimension. Two resonators with the same free spectral range $\Omega$ are coupled to yield resonant modes at $\omega_0 + \Omega n $ and $\omega_0 + \Omega n + \Omega_1$ with an integer $n$.}
    \label{fig:resonators}
\end{figure}

\subsubsection{An SSH model in the synthetic frequency dimension}
A way to realize the SSH model along the synthetic frequency dimension was proposed in Ref.~\cite{Dutt:2020:Light} and  experimentally realized in Refs.~\cite{Li:2023:Light,Pellerin:PRL2024}. We consider a variant of these setups based on a system of two coupled ring resonators, as schematically illustrated in Fig.~\ref{fig:resonators}.
The two ring resonators are considered to be identical with identical resonance frequencies. Within each resonator, the resonance frequencies are equally spaced; the spacing between them is called the {\it free spectral range} and is denoted by $\Omega$.
We denote the time-dependent amplitude of the fields in $n$-th mode of the left (right) resonator by $L_n (t)$ ($R_n (t)$), whose evolution is rule by the following equations of motion:
\begin{align}
    i\frac{d L_n (t)}{dt} &= \left( \omega_0 + \frac{\Omega_1}{2} + \Omega n \right) L_n (t) - \frac{\Omega_1}{2} R_n (t) \\
    i\frac{d R_n (t)}{dt} &= \left( \omega_0 + \frac{\Omega_1}{2} + \Omega n \right) R_n (t) - \frac{\Omega_1}{2} L_n (t)
\end{align}
Here the strength of the coupling between the two resonators is $-\Omega_1/2$ and an offset $\Omega_1/2$ is included in the single-resonator resonant frequency $\omega_0 + \Omega_1/2 + \Omega n$, so to conveniently set the frequency of the resulting supermodes.
The two families of symmetric and anti-symmetric supermodes appear as stationary states, separated by a frequency splitting of $\Omega_1$ and fully delocalized over the two resonators. The $n$-th symmetric and $n$-th anti-symmetric modes are indicate by
\begin{align}
    a_n(t) \equiv \frac{L_n (t) + R_n (t)}{\sqrt{2}}
    \\
    b_n(t) \equiv \frac{L_n (t) - R_n (t)}{\sqrt{2}};
\end{align}
their time-dependence is decoupled and obeys the equations of motion:
\begin{align}
    i\frac{d a_n (t)}{dt} &= \left( \omega_0 + \Omega n \right) a_n (t) \\
    i\frac{d b_n (t)}{dt} &= \left( \omega_0 + \Omega_1 + \Omega n \right) b_n (t).
\end{align}

As typical in synthetic frequency dimension schemes, the coupling between the supermodes is introduced by means of a temporal modulation of the frequency of the resonators. For the sake of simplicity, we focus here on the case where only the left resonator is modulated in a spatially localized way b a bichromatic modulation with components at $\Omega_1$ and $\Omega_2=\Omega-\Omega_1$ with coupling strength $2J_0$ and $2J_1$, respectively.
Given the delocalized nature of the supermodes over the two cavities, this choice of a modulation acts on both supermodes and allows to simultaneously address all transitions between pairs of supermodes.

The time-dependence of the left and right resonator fields under the modulation, including a uniform loss $\gamma$ for both resonators and a coherent pump with frequency $\omega$ and amplitude $\sqrt{2}s_\mathrm{in}$ acting only on the left resonator, is
\begin{align}
    i\frac{d L_n (t)}{dt} =& \left( \omega_0 + \frac{\Omega_1}{2} + \Omega n \right) L_n (t) - \frac{\Omega_1}{2} R_n (t)
    \notag \\
    &+ 2\left\{ J_0 \cos (\Omega_1 t) + J_1 \cos (\Omega_2 t) \right\}\sum_m L_m (t)
    \notag \\
    &- i\gamma L_n (t) + \sqrt{2}s_\mathrm{in} e^{-i\omega t}\\
    i\frac{d R_n (t)}{dt} =& \left( \omega_0 + \frac{\Omega_1}{2} + \Omega n \right) R_n (t) - \frac{\Omega_1}{2} L_n (t) - i\gamma R_n (t)
\end{align}
This translates into the following set of equations of motion for the symmetric and anti-symmetric mode amplitudes $a_n (t)$ and $b_n (t)$, 
\begin{align}
    i\frac{d a_n (t)}{dt} =& (\omega_0+\Omega n)\, a_n(t) \notag \\
    & + 2 [J_0 \cos (\Omega_1 t) + J_1 \cos (\Omega_2 t)] \sum_m (b_m+a_m) \notag \\
    &- i\gamma a_n (t) + s_{\mathrm{in}} e^{-i\omega t},
    \notag \\
    i\frac{d b_n (t)}{dt} =& (\omega_0+\Omega n + \Omega_1)b_n(t) \notag \\
    & + 2 [J_0 \cos (\Omega_1 t) + J_1 \cos (\Omega_2 t)] \sum_m (b_m+a_m) \notag  \\
    &-i \gamma b_n (t) + s_{\mathrm{in}} e^{-i\omega t}. \label{eq:wewanttosolveit}
\end{align}

Neglecting an irrelevant propagation phase around the rings and overall factors, the experimentally observable output fields $\mathrm{Out}_L(t)$ and $\mathrm{Out}_R(t)$ from left and right resonators, respectively, can be written in the form
\begin{align}
    \mathrm{Out}_L(t) &= \sum_n L_n (t) = \frac{1}{\sqrt{2}}\sum_n \left( a_n(t) + b_n (t) \right) \notag \\
    \mathrm{Out}_R(t) &= \sum_n R_n (t) = \frac{1}{\sqrt{2}}\sum_n \left( a_n(t) - b_n (t) \right).
\end{align}

We assume that both $J_0$ and $J_1$ couplings as well as the loss $\gamma$ are smaller than the mode splittings $\Omega_{1,2}$. This allows to perform a rotating wave approximation (RWA) and neglect all those coupling terms that do not resonantly couple neighboring modes.
In this regime, the band structure along the synthetic dimension is visible in the frequency direction around each resonant mode frequency, within a frequency range much smaller than the mode splittings $\Omega_{1,2}$.

We also assume that the pump frequency is in the vicinity of the central symmetric supermode (of amplitude $a_{n=0}$), so it only couples to this mode.
Note that restricting the driving to the central unit cell only is an essential assumption of the mean chiral displacement approach, and could be hardly satisfied under broadband incoherent pump schemes like the one of Ref.~\cite{StJean:2021:PRL}. The assumption of localization can be to some extent relaxed as discussed in Ref.~\cite{Derrico:PRR2020} by choosing a driving profile that is related to a localized state via a unitary transformation; fine tuning the driving field to such special classes of delocalized states is however a nontrivial task.

Putting all these assumptions together, the motion equation \eqref{eq:wewanttosolveit} reduces to a set of time-independent equations of motion for the slowly-varying amplitudes $\tilde{a}_n(t)=a_n(t)\,e^{i(\omega_0+n\Omega) t }$ and $\tilde{b}_n(t)=b_n(t)\,e^{i(\omega_0+n\Omega+\Omega_1) t }$ in the form
\begin{align}
    i\frac{d \tilde{a}_n (t)}{dt} =& J_0 \tilde{b}_n (t) + J_1 \tilde{b}_{n-1} (t)
    -i\gamma \tilde{a}_n (t) + s_{\mathrm{in}}\,\delta_{n,0} e^{-i\delta\omega t},
    \notag \\
    i\frac{d \tilde{b}_n (t)}{dt} = &  J_0 \tilde{a}_n (t) + J_1 \tilde{a}_{n+1} (t)
    -i\gamma \tilde{b}_n (t), \label{eq:wewanttosolveit2}
\end{align}
where the pump detuning from the central symmetric supermode is $\delta\omega=\omega-\omega_0$. 
We note that, while in Eq.(\ref{eq:wewanttosolveit}) the modulation couples all modes, after the RWA only the nearest neighbor couplings survive.
It is now evident that the equations of motion in Eq.~(\ref{eq:wewanttosolveit2}) exactly match the ones in Eq.~(\ref{eq:schdd}) of the abstract driven-dissipative lattice model once we identify $|\psi (t)\rangle = \begin{pmatrix} \cdots & \tilde{a}_n (t) & \tilde{b}_n (t) & \tilde{a}_{n+1} (t) & \tilde{b}_{n+1} (t) & \cdots\end{pmatrix}^T$ and the drive $|s(t)\rangle$ is set to be non-zero and equal to $s_{\mathrm{in}}\,e^{-i\,\delta\omega\, t}$ only for the A site of the central $n=0$ unit cell.

Waiting for a sufficiently long time $t\gg \gamma^{-1}$ after the source is applied, the system eventually reaches a steady state of the form $\tilde{a}_n(t)=\tilde{a}_n^{ss}\, e^{-i\delta \omega\, t}$ and $\tilde{b}_n (t)=\tilde{b}_n^{ss}\, e^{-i\,\delta \omega \, t}$, the time-independent amplitudes $\tilde{a}_n^{ss}$ and $\tilde{b}_n^{ss}$ being given by Eq.~(\ref{eq:steadygeneral}) after replacing $\omega$ by $\delta \omega$.

\begin{figure}[htbp]
    \centering
    \includegraphics[width=0.35\textwidth]{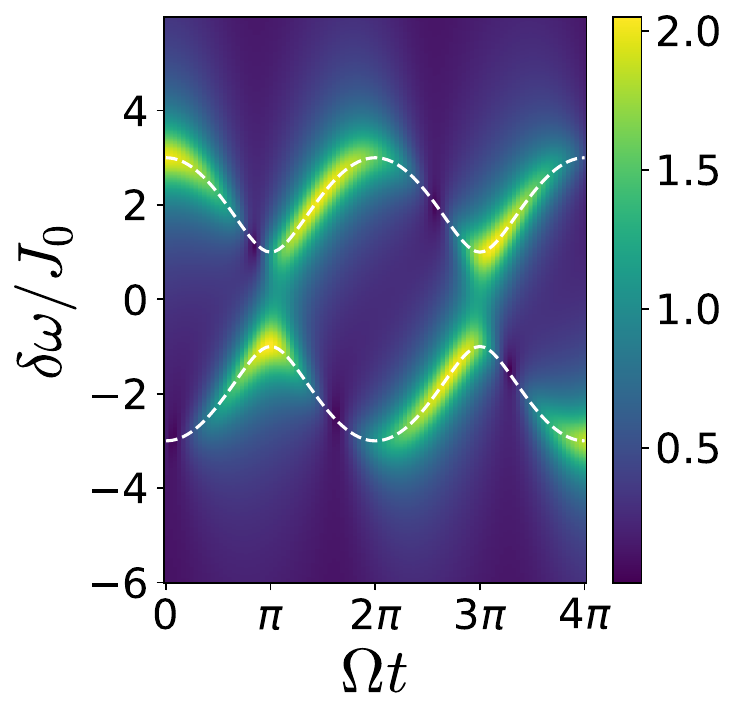}
    \caption{The output field intensity, $|\mathrm{Out}_R(t)|$. Calculation is performed following Eq.~(\ref{eq:steadyfull}) for the model with $\mathcal{W} = 1$ ($J_1= 2J_0$, $J_2=0$). We have chosen 
    $\Omega_1 = 4J_0$ and $\Omega = 20 J_0$ with $\gamma = 0.5 J_0$.
    The white dashed lines indicate the theoretical band structure of the model.}
    \label{fig:sd_simulation}
\end{figure}

The steady-state output field emitted in the waveguide coupled to the left resonator is then
\begin{multline}
    \mathrm{Out}_L(t) 
    =\sum_n \left( \tilde{a}_n^{ss}\,e^{-i(\omega+n\Omega) t } + \tilde{b}_n^{ss} \,e^{-i(\omega+n\Omega+\Omega_1) t }\right) \\ = 
    s_\mathrm{in}\,e^{-i\omega t}\, \sum_{\alpha = \pm}
    \frac{u_\alpha^A (\Omega t)^* \left\{ u_\alpha^A (\Omega t) + e^{-i\Omega_1 t} u_\alpha^B (\Omega t)\right\}}{\delta\omega - E_\alpha(\Omega t) + i\gamma}.
    \label{eq:steadyfull}
\end{multline}
where we have defined $u_\alpha^i (\Omega t)$ to be the $i=A,B$ sublattice component of the two-component vector $|u_\alpha (\Omega t)\rangle$ describing the Bloch wavefunction of the SSH lattice at momentum $\Omega t$:
as it is typical in synthetic frequency dimension schemes, the (normalized) time $\Omega t$ plays in fact the role of the momentum associated to the synthetic frequency dimension~\cite{Yjan:2018:Optica,Ehrhardt:2023:LPR}.  
Analogously, the field $\mathrm{Out}_L(t)$ that is emitted into the output waveguide coupled to the right resonator is
\begin{multline}
    \mathrm{Out}_R(t) 
    =\sum_n \left( \tilde{a}_n^{ss}\,e^{-i(\omega+n\Omega) t } - \tilde{b}_n^{ss} \,e^{-i(\omega+n\Omega+\Omega_1) t }\right)\,.
    \label{eq:steadyfull2}
\end{multline}
In an actual experiment, the complex-valued time-dependence of both output fields $\mathrm{Out}_{L,R}(t)$ can be measured with a suitable homodyning protocol, such as mixing the output fields with the pump field at frequency $\omega$. This provides slowly varying signals $S_L (t) \equiv e^{i\omega t}\mathrm{Out}_L (t)$ and $S_R (t) \equiv e^{i\omega t}\mathrm{Out}_R (t)$ that can be measured with standard electronics and provide full information on $\mathrm{Out}_{L,R}(t)$.

A colorplot summarizing the output field intensity $|\mathrm{Out}_L(t)|$ as a function of (normalized) time $\Omega t$ for different values of the detuning $\delta\omega$ is shown in Fig.\ref{fig:sd_simulation} for the steady-state of a SSH model: as typical in synthetic frequency dimension schemes~\cite{Dutt:2019:NatComm,Balcytis:2022:SciAdv}, the maximum intensity line follows the band structure of the lattice (white line). The intensity modulation that is visible on top of the band structure is typical of lattices with multiple sites per unit cell: as it was highlighted~\cite{Li:2023:Light,Pellerin:PRL2024}, its frequency is set by the in-cell frequency spacing $\Omega_1$ and, for each value of the synthetic frequency momentum $\Omega t$, its phase provides information on the relative phase of the two components of the Bloch wavefunction within the unit cell.

\subsubsection{Mean chiral displacement through spectral information}
\begin{figure}
    \centering
    \includegraphics[width=0.45\textwidth]{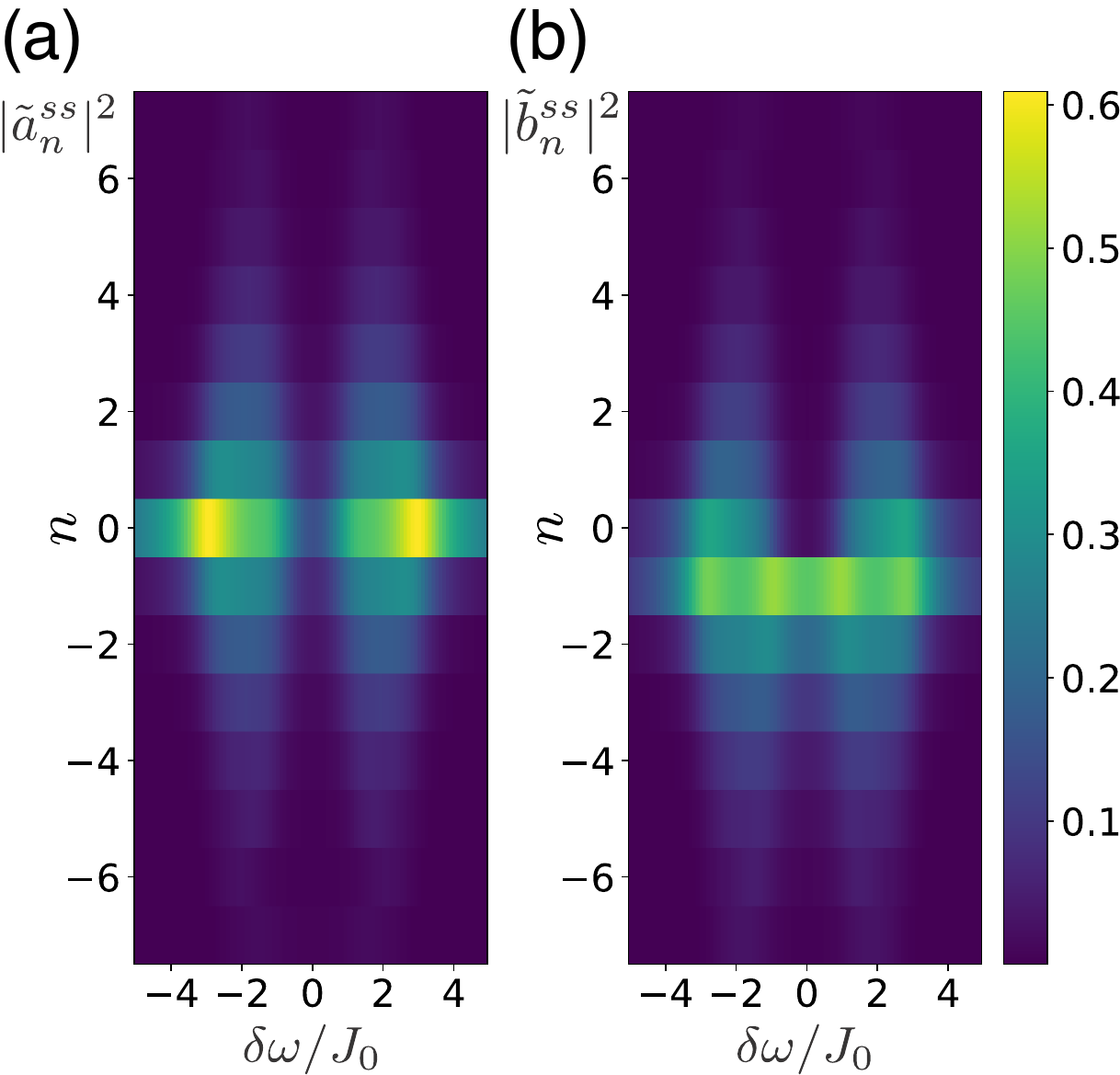}
    \caption{Colorplot of the steady-state intensities. (a) $|\tilde{a}_n^{ss}|^2$ and (b) $|\tilde{b}_n^{ss}|^2$ of the different spectral components are computed as a function of the pump frequency $\delta\omega$ and the mode (site) index $n$. We used a model with $\mathcal{W}=1$ ($J_1= 2J_0$, $J_2=0$) with $\gamma = 0.5 J_0$. Application of the mean chiral displacement formula \eqref{eq:synth_MCD} gives $2 \langle \Gamma x\rangle_\mathrm{int} = 0.97 $ to be compared to the expected value $\mathcal{W}=1$. }
    \label{fig:final}
\end{figure}

A straightforward way to extract the mean chiral displacement in the synthetic frequency dimension framework is based on measuring the intensities $|\tilde{a}_n^{ss}|^2$ and $|\tilde{b}_n^{ss}|^2$ of the different spectral components of the output signal $\mathrm{Out}_{L,R}(t)$ and then translating them into the usual definition \eqref{eq:Gammax} of mean chiral displacement to our context. This gives
\begin{align}
    \langle \Gamma x\rangle_\mathrm{int}
    =
    \frac{\int\,d\delta\omega\, \sum_n\,n\, (|\tilde{a}_n^{ss}|^2-|\tilde{b}_n^{ss}|^2)}
    {\int\,d\delta\omega\, \sum_n\, (|\tilde{a}_n^{ss}|^2+|\tilde{b}_n^{ss}|^2)}\,,
    \label{eq:synth_MCD}
\end{align}
where the integral over $\delta\omega$ has to be done on a range that is much larger than the lattice bandwidth (set by the coupling amplitudes $J_j$), but narrower than the distance $\Omega_{1,2}$ to the next site along the frequency direction.
An example of application of such procedure is illustrated in Fig.\ref{fig:final}, where we plot the intensities $|\tilde{a}_n^{ss}|^2$ and $|\tilde{b}_n^{ss}|^2$ of the different spectral components as a function of the pump frequency $\delta\omega$ for a SSH lattice with $\mathcal{W}=1$. Summing over $n$ and performing the integration along $\delta\omega$ within the plotting window gives $2\langle \Gamma x\rangle_\mathrm{int} = 0.97 $ to be compared to the expected value $\mathcal{W}=1$. This example confirms the efficiency of our proposed method to detect the lattice topology also in the case of synthetic frequency lattices. As a key advantage, the synthetic frequency framework allows to realize systems with a very large number of lattice sites~\cite{senanian2023programmable}. In this way, finite size effects are strongly suppressed and the only remaining source of discrepancy is due to photon losses.

\subsubsection{Mean chiral displacement through temporal information}
While the protocol discussed in the previous subsection is a direct translation to our context of a standard approach, one can take full advantage of the peculiarities of the synthetic dimension platform to devise an alternative protocol that does not require spectral analysis of the output signal.
The key idea is to notice that the summand in the numerator of the expression \eqref{eq:synth_MCD} for the mean chiral displacement can be rewritten under a Fourier transform as a derivative with respect to the conjugate variable associated to the synthetic frequency dimension, which is the (normalized) time $\Omega t$.

As introduced above, we consider mixing the output signals $\mathrm{Out}_{L,R}(t)$ with a monochromatic light $e^{i\omega t}$ to obtain signals $S_{L,R} (t) \equiv e^{i\omega t}\mathrm{Out}_{L,R} (t)$.
Then, for each value $\delta\omega$ of the pump frequency, one has 
\begin{align}
I_{\delta\omega}&=\int_0^{T}dt|S_{L,R}(t)|^2=\mathcal{N}\,\sum_n \left[|\tilde{a}_n^{ss}|^2 + |\tilde{b}_n^{ss}|^2\right]  \\
I^{LR}_{\delta\omega}&=\int_0^{T}dtS_L^*(t)\,S_R(t) =\mathcal{N}\,\sum_n \left[|\tilde{a}_n^{ss}|^2 - |\tilde{b}_n^{ss}|^2\right] \\
D^{LR}_{\delta\omega} &= i\,\int_0^{T}dtS_L^*(t)\,\frac{dS_R(t)}{dt} \\ &=\mathcal{N}\,\sum_n \left\{ n\Omega\,\left[|\tilde{a}_n^{ss}|^2 - |\tilde{b}_n^{ss}|^2\right] -\Omega_1\,|\tilde{b}_n^{ss}|^2\right\}\,,\nonumber
\end{align}
where $\mathcal{N}$ is a common multiplicative factor.
The upper limit $T$ of the integral, can be taken equal to $2\pi/\Omega_1$ if $\Omega$ is an integer multiple of $\Omega_1$; in general when $\Omega$ is not an integer multiple of $\Omega_1$, $T$ should be taken much larger than $2\pi/\Omega_1$ to suppress terms oscillating with frequencies $\Omega$ and $\Omega_1$.
In this limit of large $T$, by suitably combining these expressions and then integrating over $\delta\omega$, we get to an expression 
\begin{equation}
\langle \Gamma x\rangle_\mathrm{int}
    = \frac{\int\!\delta\omega\,\left[D^{LR}_{\delta\omega}+\Omega_1\,(I_{\delta\omega}-I_{\delta\omega}^{LR})/2 \right]}{\Omega\,\int\!\delta\omega\,I_{\delta\omega}}
\end{equation}
that can be used to extract the value of the integrated mean chiral displacement from measurements of the emitted signals in the two $L,R$ output waveguides.

\section{Discussion}
In this work, we have proposed a method to extract the topological winding number of one-dimensional chiral photonic lattices through the measurement of the frequency-integrated mean-chiral displacement under a coherent drive. We showed that our method can give a good estimate of the winding number in systems with realistic loss rates and with a relatively short length of lattices. Our method also works for systems with winding number greater than one, which the Zak phase, being only sensitive to the parity of the winding number, cannot discern. 

As a specific example, we have discussed the application of our method to synthetic lattices obtained via the synthetic frequency dimension scheme. Here, the ability to selectively pump a given site of the lattice with a coherent pump is a key advantage over the incoherent pumping schemes used, e.g., in polaritonic lattices~\cite{StJean:2021:PRL}, whose broadband nature would translate into a simultaneous excitation of many sites of the synthetic lattice, hindering the use of the mean chiral displacement method.
In particular, we have shown that the frequency distribution of the output light in the synthetic frequency lattice is exactly the same as the steady state distribution of a model defined in the ordinary, spatial, dimension. This equivalence between the steady state in the frequency synthetic dimension and the spatial dimension implies that the method of integrated mean chiral displacement can be used to directly detect the topological winding number of chiral Hamiltonians in the frequency synthetic dimension. 

As a future work, we plan to go beyond the simplest examples of driven-dissipative lattices studied so far where losses are assumed to be uniform and diagonal in the site basis. In these investigations, a special attention will be devoted to the study of configurations featuring addition of gain, complex non-Hermitian effects, and novel non-Hermitian topologies~\cite{Bergholtz:2021:RMP}. As a further step, it will be of great interest to extend the method to two or higher dimensions: following generalizations of the concept of mean-chiral displacement to two dimensions~\cite{Mizoguchi:2021,Wu:2021}, our formalism provides in fact a natural starting point to explore how topological invariants of higher dimensional systems, such as those in higher-order topological phases~\cite{Benalcazar:2022}, extend to driven-dissipative setups. Finally, in view of the recent development of the concept of topological bands in nonlinear systems~\cite{Jurgensen:2021,Sone:2024}, an interesting avenue for speculative investigations will be to assess how our method can be generalized to nonlinear systems.

\begin{acknowledgments}
We wish to acknowledge a continuous collaboration with Toshihiko Baba, Armandas Bal\v{c}ytis, Satoshi Iwamoto, Yasutomo Ota, F\'elix Pellerin, and Philippe St-Jean. I.C. acknowledges financial support from the PNRR-MUR project PE0000023-NQSTI project, co-funded by the European Union - NextGeneration EU, and from Provincia Autonoma di Trento (PAT), partly via the Q@TN initiative. G.V. acknowledges financial support from the Universit\`a di Trento and Advanced Institute for Materials Research at Tohoku University. T.O. acknowledges financial support from JSPS KAKENHI Grant Nos. JP20H01845 and JP24K00548, JST PRESTO Grant No. JPMJPR2353, and JST CREST Grant No. JPMJCR19T1.
\end{acknowledgments}





\bibliography{bibliography}

\begin{thebibliography}{44}%
\makeatletter
\providecommand \@ifxundefined [1]{%
 \@ifx{#1\undefined}
}%
\providecommand \@ifnum [1]{%
 \ifnum #1\expandafter \@firstoftwo
 \else \expandafter \@secondoftwo
 \fi
}%
\providecommand \@ifx [1]{%
 \ifx #1\expandafter \@firstoftwo
 \else \expandafter \@secondoftwo
 \fi
}%
\providecommand \natexlab [1]{#1}%
\providecommand \enquote  [1]{``#1''}%
\providecommand \bibnamefont  [1]{#1}%
\providecommand \bibfnamefont [1]{#1}%
\providecommand \citenamefont [1]{#1}%
\providecommand \href@noop [0]{\@secondoftwo}%
\providecommand \href [0]{\begingroup \@sanitize@url \@href}%
\providecommand \@href[1]{\@@startlink{#1}\@@href}%
\providecommand \@@href[1]{\endgroup#1\@@endlink}%
\providecommand \@sanitize@url [0]{\catcode `\\12\catcode `\$12\catcode
  `\&12\catcode `\#12\catcode `\^12\catcode `\_12\catcode `\%12\relax}%
\providecommand \@@startlink[1]{}%
\providecommand \@@endlink[0]{}%
\providecommand \url  [0]{\begingroup\@sanitize@url \@url }%
\providecommand \@url [1]{\endgroup\@href {#1}{\urlprefix }}%
\providecommand \urlprefix  [0]{URL }%
\providecommand \Eprint [0]{\href }%
\providecommand \doibase [0]{https://doi.org/}%
\providecommand \selectlanguage [0]{\@gobble}%
\providecommand \bibinfo  [0]{\@secondoftwo}%
\providecommand \bibfield  [0]{\@secondoftwo}%
\providecommand \translation [1]{[#1]}%
\providecommand \BibitemOpen [0]{}%
\providecommand \bibitemStop [0]{}%
\providecommand \bibitemNoStop [0]{.\EOS\space}%
\providecommand \EOS [0]{\spacefactor3000\relax}%
\providecommand \BibitemShut  [1]{\csname bibitem#1\endcsname}%
\let\auto@bib@innerbib\@empty
\bibitem [{\citenamefont {Lu}\ \emph {et~al.}(2014)\citenamefont {Lu},
  \citenamefont {Joannopoulos},\ and\ \citenamefont
  {Solja{\v{c}}i{\'c}}}]{Lu:2014:NatPhot}%
  \BibitemOpen
  \bibfield  {author} {\bibinfo {author} {\bibfnamefont {L.}~\bibnamefont
  {Lu}}, \bibinfo {author} {\bibfnamefont {J.~D.}\ \bibnamefont
  {Joannopoulos}},\ and\ \bibinfo {author} {\bibfnamefont {M.}~\bibnamefont
  {Solja{\v{c}}i{\'c}}},\ }\bibfield  {title} {\bibinfo {title} {Topological
  photonics},\ }\href {https://www.nature.com/articles/nphoton.2014.248}
  {\bibfield  {journal} {\bibinfo  {journal} {Nature photonics}\ }\textbf
  {\bibinfo {volume} {8}},\ \bibinfo {pages} {821} (\bibinfo {year}
  {2014})}\BibitemShut {NoStop}%
\bibitem [{\citenamefont {Ozawa}\ \emph {et~al.}(2019)\citenamefont {Ozawa},
  \citenamefont {Price}, \citenamefont {Amo}, \citenamefont {Goldman},
  \citenamefont {Hafezi}, \citenamefont {Lu}, \citenamefont {Rechtsman},
  \citenamefont {Schuster}, \citenamefont {Simon}, \citenamefont {Zilberberg},\
  and\ \citenamefont {Carusotto}}]{Ozawa:2019:RMP}%
  \BibitemOpen
  \bibfield  {author} {\bibinfo {author} {\bibfnamefont {T.}~\bibnamefont
  {Ozawa}}, \bibinfo {author} {\bibfnamefont {H.~M.}\ \bibnamefont {Price}},
  \bibinfo {author} {\bibfnamefont {A.}~\bibnamefont {Amo}}, \bibinfo {author}
  {\bibfnamefont {N.}~\bibnamefont {Goldman}}, \bibinfo {author} {\bibfnamefont
  {M.}~\bibnamefont {Hafezi}}, \bibinfo {author} {\bibfnamefont
  {L.}~\bibnamefont {Lu}}, \bibinfo {author} {\bibfnamefont {M.~C.}\
  \bibnamefont {Rechtsman}}, \bibinfo {author} {\bibfnamefont {D.}~\bibnamefont
  {Schuster}}, \bibinfo {author} {\bibfnamefont {J.}~\bibnamefont {Simon}},
  \bibinfo {author} {\bibfnamefont {O.}~\bibnamefont {Zilberberg}},\ and\
  \bibinfo {author} {\bibfnamefont {I.}~\bibnamefont {Carusotto}},\ }\bibfield
  {title} {\bibinfo {title} {Topological photonics},\ }\href
  {https://doi.org/10.1103/RevModPhys.91.015006} {\bibfield  {journal}
  {\bibinfo  {journal} {Rev. Mod. Phys.}\ }\textbf {\bibinfo {volume} {91}},\
  \bibinfo {pages} {015006} (\bibinfo {year} {2019})}\BibitemShut {NoStop}%
\bibitem [{\citenamefont {Ota}\ \emph {et~al.}(2020)\citenamefont {Ota},
  \citenamefont {Takata}, \citenamefont {Ozawa}, \citenamefont {Amo},
  \citenamefont {Jia}, \citenamefont {Kante}, \citenamefont {Notomi},
  \citenamefont {Arakawa},\ and\ \citenamefont {Iwamoto}}]{Ota:2020:Nanophot}%
  \BibitemOpen
  \bibfield  {author} {\bibinfo {author} {\bibfnamefont {Y.}~\bibnamefont
  {Ota}}, \bibinfo {author} {\bibfnamefont {K.}~\bibnamefont {Takata}},
  \bibinfo {author} {\bibfnamefont {T.}~\bibnamefont {Ozawa}}, \bibinfo
  {author} {\bibfnamefont {A.}~\bibnamefont {Amo}}, \bibinfo {author}
  {\bibfnamefont {Z.}~\bibnamefont {Jia}}, \bibinfo {author} {\bibfnamefont
  {B.}~\bibnamefont {Kante}}, \bibinfo {author} {\bibfnamefont
  {M.}~\bibnamefont {Notomi}}, \bibinfo {author} {\bibfnamefont
  {Y.}~\bibnamefont {Arakawa}},\ and\ \bibinfo {author} {\bibfnamefont
  {S.}~\bibnamefont {Iwamoto}},\ }\bibfield  {title} {\bibinfo {title} {Active
  topological photonics},\ }\href
  {https://www.degruyter.com/document/doi/10.1515/nanoph-2019-0376/html}
  {\bibfield  {journal} {\bibinfo  {journal} {Nanophotonics}\ }\textbf
  {\bibinfo {volume} {9}},\ \bibinfo {pages} {547} (\bibinfo {year}
  {2020})}\BibitemShut {NoStop}%
\bibitem [{\citenamefont {Segev}\ and\ \citenamefont
  {Bandres}(2020)}]{Segev:2020:Nanophot}%
  \BibitemOpen
  \bibfield  {author} {\bibinfo {author} {\bibfnamefont {M.}~\bibnamefont
  {Segev}}\ and\ \bibinfo {author} {\bibfnamefont {M.~A.}\ \bibnamefont
  {Bandres}},\ }\bibfield  {title} {\bibinfo {title} {Topological photonics:
  Where do we go from here?},\ }\href
  {https://www.degruyter.com/document/doi/10.1515/nanoph-2020-0441/html}
  {\bibfield  {journal} {\bibinfo  {journal} {Nanophotonics}\ }\textbf
  {\bibinfo {volume} {10}},\ \bibinfo {pages} {425} (\bibinfo {year}
  {2020})}\BibitemShut {NoStop}%
\bibitem [{\citenamefont {Chen}\ and\ \citenamefont
  {Segev}(2021)}]{Chen:2021:Elight}%
  \BibitemOpen
  \bibfield  {author} {\bibinfo {author} {\bibfnamefont {Z.}~\bibnamefont
  {Chen}}\ and\ \bibinfo {author} {\bibfnamefont {M.}~\bibnamefont {Segev}},\
  }\bibfield  {title} {\bibinfo {title} {Highlighting photonics: looking into
  the next decade},\ }\href
  {https://elight.springeropen.com/articles/10.1186/s43593-021-00002-y}
  {\bibfield  {journal} {\bibinfo  {journal} {ELight}\ }\textbf {\bibinfo
  {volume} {1}},\ \bibinfo {pages} {2} (\bibinfo {year} {2021})}\BibitemShut
  {NoStop}%
\bibitem [{\citenamefont {Price}\ \emph {et~al.}(2022)\citenamefont {Price},
  \citenamefont {Chong}, \citenamefont {Khanikaev}, \citenamefont {Schomerus},
  \citenamefont {Maczewsky}, \citenamefont {Kremer}, \citenamefont {Heinrich},
  \citenamefont {Szameit}, \citenamefont {Zilberberg}, \citenamefont {Yang}
  \emph {et~al.}}]{Price:2022:Roadmap}%
  \BibitemOpen
  \bibfield  {author} {\bibinfo {author} {\bibfnamefont {H.}~\bibnamefont
  {Price}}, \bibinfo {author} {\bibfnamefont {Y.}~\bibnamefont {Chong}},
  \bibinfo {author} {\bibfnamefont {A.}~\bibnamefont {Khanikaev}}, \bibinfo
  {author} {\bibfnamefont {H.}~\bibnamefont {Schomerus}}, \bibinfo {author}
  {\bibfnamefont {L.~J.}\ \bibnamefont {Maczewsky}}, \bibinfo {author}
  {\bibfnamefont {M.}~\bibnamefont {Kremer}}, \bibinfo {author} {\bibfnamefont
  {M.}~\bibnamefont {Heinrich}}, \bibinfo {author} {\bibfnamefont
  {A.}~\bibnamefont {Szameit}}, \bibinfo {author} {\bibfnamefont
  {O.}~\bibnamefont {Zilberberg}}, \bibinfo {author} {\bibfnamefont
  {Y.}~\bibnamefont {Yang}}, \emph {et~al.},\ }\bibfield  {title} {\bibinfo
  {title} {Roadmap on topological photonics},\ }\href
  {https://iopscience.iop.org/article/10.1088/2515-7647/ac4ee4} {\bibfield
  {journal} {\bibinfo  {journal} {Journal of Physics: Photonics}\ }\textbf
  {\bibinfo {volume} {4}},\ \bibinfo {pages} {032501} (\bibinfo {year}
  {2022})}\BibitemShut {NoStop}%
\bibitem [{\citenamefont {Zhang}\ \emph {et~al.}(2023)\citenamefont {Zhang},
  \citenamefont {Zangeneh-Nejad}, \citenamefont {Chen}, \citenamefont {Lu},\
  and\ \citenamefont {Christensen}}]{Zhang:2023:Nature}%
  \BibitemOpen
  \bibfield  {author} {\bibinfo {author} {\bibfnamefont {X.}~\bibnamefont
  {Zhang}}, \bibinfo {author} {\bibfnamefont {F.}~\bibnamefont
  {Zangeneh-Nejad}}, \bibinfo {author} {\bibfnamefont {Z.-G.}\ \bibnamefont
  {Chen}}, \bibinfo {author} {\bibfnamefont {M.-H.}\ \bibnamefont {Lu}},\ and\
  \bibinfo {author} {\bibfnamefont {J.}~\bibnamefont {Christensen}},\
  }\bibfield  {title} {\bibinfo {title} {A second wave of topological phenomena
  in photonics and acoustics},\ }\href
  {https://www.nature.com/articles/s41586-023-06163-9} {\bibfield  {journal}
  {\bibinfo  {journal} {Nature}\ }\textbf {\bibinfo {volume} {618}},\ \bibinfo
  {pages} {687} (\bibinfo {year} {2023})}\BibitemShut {NoStop}%
\bibitem [{\citenamefont {Asb{\'o}th}\ \emph {et~al.}(2016)\citenamefont
  {Asb{\'o}th}, \citenamefont {Oroszl{\'a}ny},\ and\ \citenamefont
  {P{\'a}lyi}}]{Asboth:2016:Lecture}%
  \BibitemOpen
  \bibfield  {author} {\bibinfo {author} {\bibfnamefont {J.~K.}\ \bibnamefont
  {Asb{\'o}th}}, \bibinfo {author} {\bibfnamefont {L.}~\bibnamefont
  {Oroszl{\'a}ny}},\ and\ \bibinfo {author} {\bibfnamefont {A.}~\bibnamefont
  {P{\'a}lyi}},\ }\bibfield  {title} {\bibinfo {title} {A short course on
  topological insulators},\ }\href
  {https://link.springer.com/book/10.1007/978-3-319-25607-8} {\bibfield
  {journal} {\bibinfo  {journal} {Lecture notes in physics}\ }\textbf {\bibinfo
  {volume} {919}},\ \bibinfo {pages} {166} (\bibinfo {year}
  {2016})}\BibitemShut {NoStop}%
\bibitem [{\citenamefont {Su}\ \emph {et~al.}(1979)\citenamefont {Su},
  \citenamefont {Schrieffer},\ and\ \citenamefont {Heeger}}]{Su:1979:PRL}%
  \BibitemOpen
  \bibfield  {author} {\bibinfo {author} {\bibfnamefont {W.~P.}\ \bibnamefont
  {Su}}, \bibinfo {author} {\bibfnamefont {J.~R.}\ \bibnamefont {Schrieffer}},\
  and\ \bibinfo {author} {\bibfnamefont {A.~J.}\ \bibnamefont {Heeger}},\
  }\bibfield  {title} {\bibinfo {title} {Solitons in polyacetylene},\ }\href
  {https://doi.org/10.1103/PhysRevLett.42.1698} {\bibfield  {journal} {\bibinfo
   {journal} {Phys. Rev. Lett.}\ }\textbf {\bibinfo {volume} {42}},\ \bibinfo
  {pages} {1698} (\bibinfo {year} {1979})}\BibitemShut {NoStop}%
\bibitem [{\citenamefont {Chiu}\ \emph {et~al.}(2016)\citenamefont {Chiu},
  \citenamefont {Teo}, \citenamefont {Schnyder},\ and\ \citenamefont
  {Ryu}}]{Chiu:2016:RMP}%
  \BibitemOpen
  \bibfield  {author} {\bibinfo {author} {\bibfnamefont {C.-K.}\ \bibnamefont
  {Chiu}}, \bibinfo {author} {\bibfnamefont {J.~C.~Y.}\ \bibnamefont {Teo}},
  \bibinfo {author} {\bibfnamefont {A.~P.}\ \bibnamefont {Schnyder}},\ and\
  \bibinfo {author} {\bibfnamefont {S.}~\bibnamefont {Ryu}},\ }\bibfield
  {title} {\bibinfo {title} {Classification of topological quantum matter with
  symmetries},\ }\href {https://doi.org/10.1103/RevModPhys.88.035005}
  {\bibfield  {journal} {\bibinfo  {journal} {Rev. Mod. Phys.}\ }\textbf
  {\bibinfo {volume} {88}},\ \bibinfo {pages} {035005} (\bibinfo {year}
  {2016})}\BibitemShut {NoStop}%
\bibitem [{\citenamefont {Ozawa}\ and\ \citenamefont
  {Carusotto}(2014)}]{Ozawa:2014:PRL}%
  \BibitemOpen
  \bibfield  {author} {\bibinfo {author} {\bibfnamefont {T.}~\bibnamefont
  {Ozawa}}\ and\ \bibinfo {author} {\bibfnamefont {I.}~\bibnamefont
  {Carusotto}},\ }\bibfield  {title} {\bibinfo {title} {Anomalous and quantum
  {H}all effects in lossy photonic lattices},\ }\href
  {https://doi.org/10.1103/PhysRevLett.112.133902} {\bibfield  {journal}
  {\bibinfo  {journal} {Phys. Rev. Lett.}\ }\textbf {\bibinfo {volume} {112}},\
  \bibinfo {pages} {133902} (\bibinfo {year} {2014})}\BibitemShut {NoStop}%
\bibitem [{\citenamefont {Bardyn}\ \emph {et~al.}(2014)\citenamefont {Bardyn},
  \citenamefont {Huber},\ and\ \citenamefont
  {Zilberberg}}]{bardyn2014measuring}%
  \BibitemOpen
  \bibfield  {author} {\bibinfo {author} {\bibfnamefont {C.-E.}\ \bibnamefont
  {Bardyn}}, \bibinfo {author} {\bibfnamefont {S.~D.}\ \bibnamefont {Huber}},\
  and\ \bibinfo {author} {\bibfnamefont {O.}~\bibnamefont {Zilberberg}},\
  }\bibfield  {title} {\bibinfo {title} {Measuring topological invariants in
  small photonic lattices},\ }\href
  {https://iopscience.iop.org/article/10.1088/1367-2630/16/12/123013/meta}
  {\bibfield  {journal} {\bibinfo  {journal} {New Journal of Physics}\ }\textbf
  {\bibinfo {volume} {16}},\ \bibinfo {pages} {123013} (\bibinfo {year}
  {2014})}\BibitemShut {NoStop}%
\bibitem [{\citenamefont {Wimmer}\ \emph {et~al.}(2017)\citenamefont {Wimmer},
  \citenamefont {Price}, \citenamefont {Carusotto},\ and\ \citenamefont
  {Peschel}}]{Wimmer:2017:NatPhys}%
  \BibitemOpen
  \bibfield  {author} {\bibinfo {author} {\bibfnamefont {M.}~\bibnamefont
  {Wimmer}}, \bibinfo {author} {\bibfnamefont {H.~M.}\ \bibnamefont {Price}},
  \bibinfo {author} {\bibfnamefont {I.}~\bibnamefont {Carusotto}},\ and\
  \bibinfo {author} {\bibfnamefont {U.}~\bibnamefont {Peschel}},\ }\bibfield
  {title} {\bibinfo {title} {Experimental measurement of the {B}erry curvature
  from anomalous transport},\ }\href
  {https://www.nature.com/articles/nphys4050} {\bibfield  {journal} {\bibinfo
  {journal} {Nature Physics}\ }\textbf {\bibinfo {volume} {13}},\ \bibinfo
  {pages} {545} (\bibinfo {year} {2017})}\BibitemShut {NoStop}%
\bibitem [{\citenamefont {Gianfrate}\ \emph {et~al.}(2020)\citenamefont
  {Gianfrate}, \citenamefont {Bleu}, \citenamefont {Dominici}, \citenamefont
  {Ardizzone}, \citenamefont {De~Giorgi}, \citenamefont {Ballarini},
  \citenamefont {Lerario}, \citenamefont {West}, \citenamefont {Pfeiffer},
  \citenamefont {Solnyshkov} \emph {et~al.}}]{Gianfrate:2020:Nature}%
  \BibitemOpen
  \bibfield  {author} {\bibinfo {author} {\bibfnamefont {A.}~\bibnamefont
  {Gianfrate}}, \bibinfo {author} {\bibfnamefont {O.}~\bibnamefont {Bleu}},
  \bibinfo {author} {\bibfnamefont {L.}~\bibnamefont {Dominici}}, \bibinfo
  {author} {\bibfnamefont {V.}~\bibnamefont {Ardizzone}}, \bibinfo {author}
  {\bibfnamefont {M.}~\bibnamefont {De~Giorgi}}, \bibinfo {author}
  {\bibfnamefont {D.}~\bibnamefont {Ballarini}}, \bibinfo {author}
  {\bibfnamefont {G.}~\bibnamefont {Lerario}}, \bibinfo {author} {\bibfnamefont
  {K.}~\bibnamefont {West}}, \bibinfo {author} {\bibfnamefont {L.}~\bibnamefont
  {Pfeiffer}}, \bibinfo {author} {\bibfnamefont {D.}~\bibnamefont
  {Solnyshkov}}, \emph {et~al.},\ }\bibfield  {title} {\bibinfo {title}
  {Measurement of the quantum geometric tensor and of the anomalous {H}all
  drift},\ }\href {https://www.nature.com/articles/s41586-020-1989-2}
  {\bibfield  {journal} {\bibinfo  {journal} {Nature}\ }\textbf {\bibinfo
  {volume} {578}},\ \bibinfo {pages} {381} (\bibinfo {year}
  {2020})}\BibitemShut {NoStop}%
\bibitem [{\citenamefont {Cardano}\ \emph {et~al.}(2017)\citenamefont
  {Cardano}, \citenamefont {D’Errico}, \citenamefont {Dauphin}, \citenamefont
  {Maffei}, \citenamefont {Piccirillo}, \citenamefont {de~Lisio}, \citenamefont
  {De~Filippis}, \citenamefont {Cataudella}, \citenamefont {Santamato},
  \citenamefont {Marrucci} \emph {et~al.}}]{Cardano:2017:NatComm}%
  \BibitemOpen
  \bibfield  {author} {\bibinfo {author} {\bibfnamefont {F.}~\bibnamefont
  {Cardano}}, \bibinfo {author} {\bibfnamefont {A.}~\bibnamefont {D’Errico}},
  \bibinfo {author} {\bibfnamefont {A.}~\bibnamefont {Dauphin}}, \bibinfo
  {author} {\bibfnamefont {M.}~\bibnamefont {Maffei}}, \bibinfo {author}
  {\bibfnamefont {B.}~\bibnamefont {Piccirillo}}, \bibinfo {author}
  {\bibfnamefont {C.}~\bibnamefont {de~Lisio}}, \bibinfo {author}
  {\bibfnamefont {G.}~\bibnamefont {De~Filippis}}, \bibinfo {author}
  {\bibfnamefont {V.}~\bibnamefont {Cataudella}}, \bibinfo {author}
  {\bibfnamefont {E.}~\bibnamefont {Santamato}}, \bibinfo {author}
  {\bibfnamefont {L.}~\bibnamefont {Marrucci}}, \emph {et~al.},\ }\bibfield
  {title} {\bibinfo {title} {Detection of {Z}ak phases and topological
  invariants in a chiral quantum walk of twisted photons},\ }\href
  {https://www.nature.com/articles/ncomms15516} {\bibfield  {journal} {\bibinfo
   {journal} {Nature communications}\ }\textbf {\bibinfo {volume} {8}},\
  \bibinfo {pages} {15516} (\bibinfo {year} {2017})}\BibitemShut {NoStop}%
\bibitem [{\citenamefont {Maffei}\ \emph {et~al.}(2018)\citenamefont {Maffei},
  \citenamefont {Dauphin}, \citenamefont {Cardano}, \citenamefont
  {Lewenstein},\ and\ \citenamefont {Massignan}}]{Maffei:2018:NJP}%
  \BibitemOpen
  \bibfield  {author} {\bibinfo {author} {\bibfnamefont {M.}~\bibnamefont
  {Maffei}}, \bibinfo {author} {\bibfnamefont {A.}~\bibnamefont {Dauphin}},
  \bibinfo {author} {\bibfnamefont {F.}~\bibnamefont {Cardano}}, \bibinfo
  {author} {\bibfnamefont {M.}~\bibnamefont {Lewenstein}},\ and\ \bibinfo
  {author} {\bibfnamefont {P.}~\bibnamefont {Massignan}},\ }\bibfield  {title}
  {\bibinfo {title} {Topological characterization of chiral models through
  their long time dynamics},\ }\href
  {https://iopscience.iop.org/article/10.1088/1367-2630/aa9d4c} {\bibfield
  {journal} {\bibinfo  {journal} {New Journal of Physics}\ }\textbf {\bibinfo
  {volume} {20}},\ \bibinfo {pages} {013023} (\bibinfo {year}
  {2018})}\BibitemShut {NoStop}%
\bibitem [{\citenamefont {Longhi}(2018)}]{Longhi:2018:OptLett}%
  \BibitemOpen
  \bibfield  {author} {\bibinfo {author} {\bibfnamefont {S.}~\bibnamefont
  {Longhi}},\ }\bibfield  {title} {\bibinfo {title} {Probing one-dimensional
  topological phases in waveguide lattices with broken chiral symmetry},\
  }\href {https://opg.optica.org/ol/abstract.cfm?uri=ol-43-19-4639} {\bibfield
  {journal} {\bibinfo  {journal} {Optics Letters}\ }\textbf {\bibinfo {volume}
  {43}},\ \bibinfo {pages} {4639} (\bibinfo {year} {2018})}\BibitemShut
  {NoStop}%
\bibitem [{\citenamefont {Jiao}\ \emph {et~al.}(2021)\citenamefont {Jiao},
  \citenamefont {Longhi}, \citenamefont {Wang}, \citenamefont {Gao},
  \citenamefont {Zhou}, \citenamefont {Wang}, \citenamefont {Fu}, \citenamefont
  {Wang}, \citenamefont {Ren}, \citenamefont {Qiao},\ and\ \citenamefont
  {Jin}}]{Jiao:2021:PRL}%
  \BibitemOpen
  \bibfield  {author} {\bibinfo {author} {\bibfnamefont {Z.-Q.}\ \bibnamefont
  {Jiao}}, \bibinfo {author} {\bibfnamefont {S.}~\bibnamefont {Longhi}},
  \bibinfo {author} {\bibfnamefont {X.-W.}\ \bibnamefont {Wang}}, \bibinfo
  {author} {\bibfnamefont {J.}~\bibnamefont {Gao}}, \bibinfo {author}
  {\bibfnamefont {W.-H.}\ \bibnamefont {Zhou}}, \bibinfo {author}
  {\bibfnamefont {Y.}~\bibnamefont {Wang}}, \bibinfo {author} {\bibfnamefont
  {Y.-X.}\ \bibnamefont {Fu}}, \bibinfo {author} {\bibfnamefont
  {L.}~\bibnamefont {Wang}}, \bibinfo {author} {\bibfnamefont {R.-J.}\
  \bibnamefont {Ren}}, \bibinfo {author} {\bibfnamefont {L.-F.}\ \bibnamefont
  {Qiao}},\ and\ \bibinfo {author} {\bibfnamefont {X.-M.}\ \bibnamefont
  {Jin}},\ }\bibfield  {title} {\bibinfo {title} {Experimentally detecting
  quantized {Z}ak phases without chiral symmetry in photonic lattices},\ }\href
  {https://doi.org/10.1103/PhysRevLett.127.147401} {\bibfield  {journal}
  {\bibinfo  {journal} {Phys. Rev. Lett.}\ }\textbf {\bibinfo {volume} {127}},\
  \bibinfo {pages} {147401} (\bibinfo {year} {2021})}\BibitemShut {NoStop}%
\bibitem [{\citenamefont {C{\'a}ceres-Aravena}\ \emph
  {et~al.}(2023)\citenamefont {C{\'a}ceres-Aravena}, \citenamefont {Real},
  \citenamefont {Guzm{\'a}n-Silva}, \citenamefont {Vildoso}, \citenamefont
  {Salinas}, \citenamefont {Amo}, \citenamefont {Ozawa},\ and\ \citenamefont
  {Vicencio}}]{caceres2023edge}%
  \BibitemOpen
  \bibfield  {author} {\bibinfo {author} {\bibfnamefont {G.}~\bibnamefont
  {C{\'a}ceres-Aravena}}, \bibinfo {author} {\bibfnamefont {B.}~\bibnamefont
  {Real}}, \bibinfo {author} {\bibfnamefont {D.}~\bibnamefont
  {Guzm{\'a}n-Silva}}, \bibinfo {author} {\bibfnamefont {P.}~\bibnamefont
  {Vildoso}}, \bibinfo {author} {\bibfnamefont {I.}~\bibnamefont {Salinas}},
  \bibinfo {author} {\bibfnamefont {A.}~\bibnamefont {Amo}}, \bibinfo {author}
  {\bibfnamefont {T.}~\bibnamefont {Ozawa}},\ and\ \bibinfo {author}
  {\bibfnamefont {R.~A.}\ \bibnamefont {Vicencio}},\ }\bibfield  {title}
  {\bibinfo {title} {Edge-to-edge topological spectral transfer in diamond
  photonic lattices},\ }\href
  {https://pubs.aip.org/aip/app/article/8/8/080801/2904988/Edge-to-edge-topological-spectral-transfer-in}
  {\bibfield  {journal} {\bibinfo  {journal} {APL Photonics}\ }\textbf
  {\bibinfo {volume} {8}} (\bibinfo {year} {2023})}\BibitemShut {NoStop}%
\bibitem [{\citenamefont {St-Jean}\ \emph {et~al.}(2021)\citenamefont
  {St-Jean}, \citenamefont {Dauphin}, \citenamefont {Massignan}, \citenamefont
  {Real}, \citenamefont {Jamadi}, \citenamefont {Milicevic}, \citenamefont
  {Lema\^{\i}tre}, \citenamefont {Harouri}, \citenamefont {Le~Gratiet},
  \citenamefont {Sagnes}, \citenamefont {Ravets}, \citenamefont {Bloch},\ and\
  \citenamefont {Amo}}]{StJean:2021:PRL}%
  \BibitemOpen
  \bibfield  {author} {\bibinfo {author} {\bibfnamefont {P.}~\bibnamefont
  {St-Jean}}, \bibinfo {author} {\bibfnamefont {A.}~\bibnamefont {Dauphin}},
  \bibinfo {author} {\bibfnamefont {P.}~\bibnamefont {Massignan}}, \bibinfo
  {author} {\bibfnamefont {B.}~\bibnamefont {Real}}, \bibinfo {author}
  {\bibfnamefont {O.}~\bibnamefont {Jamadi}}, \bibinfo {author} {\bibfnamefont
  {M.}~\bibnamefont {Milicevic}}, \bibinfo {author} {\bibfnamefont
  {A.}~\bibnamefont {Lema\^{\i}tre}}, \bibinfo {author} {\bibfnamefont
  {A.}~\bibnamefont {Harouri}}, \bibinfo {author} {\bibfnamefont
  {L.}~\bibnamefont {Le~Gratiet}}, \bibinfo {author} {\bibfnamefont
  {I.}~\bibnamefont {Sagnes}}, \bibinfo {author} {\bibfnamefont
  {S.}~\bibnamefont {Ravets}}, \bibinfo {author} {\bibfnamefont
  {J.}~\bibnamefont {Bloch}},\ and\ \bibinfo {author} {\bibfnamefont
  {A.}~\bibnamefont {Amo}},\ }\bibfield  {title} {\bibinfo {title} {Measuring
  topological invariants in a polaritonic analog of graphene},\ }\href
  {https://doi.org/10.1103/PhysRevLett.126.127403} {\bibfield  {journal}
  {\bibinfo  {journal} {Phys. Rev. Lett.}\ }\textbf {\bibinfo {volume} {126}},\
  \bibinfo {pages} {127403} (\bibinfo {year} {2021})}\BibitemShut {NoStop}%
\bibitem [{\citenamefont {Ozawa}\ \emph {et~al.}(2016)\citenamefont {Ozawa},
  \citenamefont {Price}, \citenamefont {Goldman}, \citenamefont {Zilberberg},\
  and\ \citenamefont {Carusotto}}]{Ozawa:2016:PRA}%
  \BibitemOpen
  \bibfield  {author} {\bibinfo {author} {\bibfnamefont {T.}~\bibnamefont
  {Ozawa}}, \bibinfo {author} {\bibfnamefont {H.~M.}\ \bibnamefont {Price}},
  \bibinfo {author} {\bibfnamefont {N.}~\bibnamefont {Goldman}}, \bibinfo
  {author} {\bibfnamefont {O.}~\bibnamefont {Zilberberg}},\ and\ \bibinfo
  {author} {\bibfnamefont {I.}~\bibnamefont {Carusotto}},\ }\bibfield  {title}
  {\bibinfo {title} {Synthetic dimensions in integrated photonics: From optical
  isolation to four-dimensional quantum {H}all physics},\ }\href
  {https://doi.org/10.1103/PhysRevA.93.043827} {\bibfield  {journal} {\bibinfo
  {journal} {Phys. Rev. A}\ }\textbf {\bibinfo {volume} {93}},\ \bibinfo
  {pages} {043827} (\bibinfo {year} {2016})}\BibitemShut {NoStop}%
\bibitem [{\citenamefont {Yuan}\ \emph {et~al.}(2016)\citenamefont {Yuan},
  \citenamefont {Shi},\ and\ \citenamefont {Fan}}]{Yuan:2016:OptLett}%
  \BibitemOpen
  \bibfield  {author} {\bibinfo {author} {\bibfnamefont {L.}~\bibnamefont
  {Yuan}}, \bibinfo {author} {\bibfnamefont {Y.}~\bibnamefont {Shi}},\ and\
  \bibinfo {author} {\bibfnamefont {S.}~\bibnamefont {Fan}},\ }\bibfield
  {title} {\bibinfo {title} {Photonic gauge potential in a system with a
  synthetic frequency dimension},\ }\href
  {https://opg.optica.org/ol/abstract.cfm?uri=ol-41-4-741} {\bibfield
  {journal} {\bibinfo  {journal} {Optics letters}\ }\textbf {\bibinfo {volume}
  {41}},\ \bibinfo {pages} {741} (\bibinfo {year} {2016})}\BibitemShut
  {NoStop}%
\bibitem [{\citenamefont {Yuan}\ \emph {et~al.}(2018)\citenamefont {Yuan},
  \citenamefont {Lin}, \citenamefont {Xiao},\ and\ \citenamefont
  {Fan}}]{Yjan:2018:Optica}%
  \BibitemOpen
  \bibfield  {author} {\bibinfo {author} {\bibfnamefont {L.}~\bibnamefont
  {Yuan}}, \bibinfo {author} {\bibfnamefont {Q.}~\bibnamefont {Lin}}, \bibinfo
  {author} {\bibfnamefont {M.}~\bibnamefont {Xiao}},\ and\ \bibinfo {author}
  {\bibfnamefont {S.}~\bibnamefont {Fan}},\ }\bibfield  {title} {\bibinfo
  {title} {Synthetic dimension in photonics},\ }\href
  {https://opg.optica.org/optica/fulltext.cfm?uri=optica-5-11-1396&id=400075}
  {\bibfield  {journal} {\bibinfo  {journal} {Optica}\ }\textbf {\bibinfo
  {volume} {5}},\ \bibinfo {pages} {1396} (\bibinfo {year} {2018})}\BibitemShut
  {NoStop}%
\bibitem [{\citenamefont {Ozawa}\ and\ \citenamefont
  {Price}(2019)}]{Ozawa:2019:NatRevPhys}%
  \BibitemOpen
  \bibfield  {author} {\bibinfo {author} {\bibfnamefont {T.}~\bibnamefont
  {Ozawa}}\ and\ \bibinfo {author} {\bibfnamefont {H.~M.}\ \bibnamefont
  {Price}},\ }\bibfield  {title} {\bibinfo {title} {Topological quantum matter
  in synthetic dimensions},\ }\href
  {https://www.nature.com/articles/s42254-019-0045-3} {\bibfield  {journal}
  {\bibinfo  {journal} {Nature Reviews Physics}\ }\textbf {\bibinfo {volume}
  {1}},\ \bibinfo {pages} {349} (\bibinfo {year} {2019})}\BibitemShut {NoStop}%
\bibitem [{\citenamefont {Dutt}\ \emph {et~al.}(2019)\citenamefont {Dutt},
  \citenamefont {Minkov}, \citenamefont {Lin}, \citenamefont {Yuan},
  \citenamefont {Miller},\ and\ \citenamefont {Fan}}]{Dutt:2019:NatComm}%
  \BibitemOpen
  \bibfield  {author} {\bibinfo {author} {\bibfnamefont {A.}~\bibnamefont
  {Dutt}}, \bibinfo {author} {\bibfnamefont {M.}~\bibnamefont {Minkov}},
  \bibinfo {author} {\bibfnamefont {Q.}~\bibnamefont {Lin}}, \bibinfo {author}
  {\bibfnamefont {L.}~\bibnamefont {Yuan}}, \bibinfo {author} {\bibfnamefont
  {D.~A.}\ \bibnamefont {Miller}},\ and\ \bibinfo {author} {\bibfnamefont
  {S.}~\bibnamefont {Fan}},\ }\bibfield  {title} {\bibinfo {title}
  {Experimental band structure spectroscopy along a synthetic dimension},\
  }\href {https://www.nature.com/articles/s41467-019-11117-9} {\bibfield
  {journal} {\bibinfo  {journal} {Nature communications}\ }\textbf {\bibinfo
  {volume} {10}},\ \bibinfo {pages} {3122} (\bibinfo {year}
  {2019})}\BibitemShut {NoStop}%
\bibitem [{\citenamefont {Lustig}\ \emph {et~al.}(2019)\citenamefont {Lustig},
  \citenamefont {Weimann}, \citenamefont {Plotnik}, \citenamefont {Lumer},
  \citenamefont {Bandres}, \citenamefont {Szameit},\ and\ \citenamefont
  {Segev}}]{Lustig:2019:Nature}%
  \BibitemOpen
  \bibfield  {author} {\bibinfo {author} {\bibfnamefont {E.}~\bibnamefont
  {Lustig}}, \bibinfo {author} {\bibfnamefont {S.}~\bibnamefont {Weimann}},
  \bibinfo {author} {\bibfnamefont {Y.}~\bibnamefont {Plotnik}}, \bibinfo
  {author} {\bibfnamefont {Y.}~\bibnamefont {Lumer}}, \bibinfo {author}
  {\bibfnamefont {M.~A.}\ \bibnamefont {Bandres}}, \bibinfo {author}
  {\bibfnamefont {A.}~\bibnamefont {Szameit}},\ and\ \bibinfo {author}
  {\bibfnamefont {M.}~\bibnamefont {Segev}},\ }\bibfield  {title} {\bibinfo
  {title} {Photonic topological insulator in synthetic dimensions},\ }\href
  {https://www.nature.com/articles/s41586-019-0943-7} {\bibfield  {journal}
  {\bibinfo  {journal} {Nature}\ }\textbf {\bibinfo {volume} {567}},\ \bibinfo
  {pages} {356} (\bibinfo {year} {2019})}\BibitemShut {NoStop}%
\bibitem [{\citenamefont {Dutt}\ \emph
  {et~al.}(2020{\natexlab{a}})\citenamefont {Dutt}, \citenamefont {Minkov},
  \citenamefont {Williamson},\ and\ \citenamefont {Fan}}]{Dutt:2020:Light}%
  \BibitemOpen
  \bibfield  {author} {\bibinfo {author} {\bibfnamefont {A.}~\bibnamefont
  {Dutt}}, \bibinfo {author} {\bibfnamefont {M.}~\bibnamefont {Minkov}},
  \bibinfo {author} {\bibfnamefont {I.~A.}\ \bibnamefont {Williamson}},\ and\
  \bibinfo {author} {\bibfnamefont {S.}~\bibnamefont {Fan}},\ }\bibfield
  {title} {\bibinfo {title} {Higher-order topological insulators in synthetic
  dimensions},\ }\href {https://www.nature.com/articles/s41377-020-0334-8}
  {\bibfield  {journal} {\bibinfo  {journal} {Light: Science \& Applications}\
  }\textbf {\bibinfo {volume} {9}},\ \bibinfo {pages} {131} (\bibinfo {year}
  {2020}{\natexlab{a}})}\BibitemShut {NoStop}%
\bibitem [{\citenamefont {Dutt}\ \emph
  {et~al.}(2020{\natexlab{b}})\citenamefont {Dutt}, \citenamefont {Lin},
  \citenamefont {Yuan}, \citenamefont {Minkov}, \citenamefont {Xiao},\ and\
  \citenamefont {Fan}}]{Dutt:2020:Science}%
  \BibitemOpen
  \bibfield  {author} {\bibinfo {author} {\bibfnamefont {A.}~\bibnamefont
  {Dutt}}, \bibinfo {author} {\bibfnamefont {Q.}~\bibnamefont {Lin}}, \bibinfo
  {author} {\bibfnamefont {L.}~\bibnamefont {Yuan}}, \bibinfo {author}
  {\bibfnamefont {M.}~\bibnamefont {Minkov}}, \bibinfo {author} {\bibfnamefont
  {M.}~\bibnamefont {Xiao}},\ and\ \bibinfo {author} {\bibfnamefont
  {S.}~\bibnamefont {Fan}},\ }\bibfield  {title} {\bibinfo {title} {A single
  photonic cavity with two independent physical synthetic dimensions},\ }\href
  {https://www.science.org/doi/10.1126/science.aaz3071} {\bibfield  {journal}
  {\bibinfo  {journal} {Science}\ }\textbf {\bibinfo {volume} {367}},\ \bibinfo
  {pages} {59} (\bibinfo {year} {2020}{\natexlab{b}})}\BibitemShut {NoStop}%
\bibitem [{\citenamefont {Wang}\ \emph {et~al.}(2021)\citenamefont {Wang},
  \citenamefont {Dutt}, \citenamefont {Yang}, \citenamefont {Wojcik},
  \citenamefont {Vu{\v{c}}kovi{\'c}},\ and\ \citenamefont
  {Fan}}]{Wang:2021:Science}%
  \BibitemOpen
  \bibfield  {author} {\bibinfo {author} {\bibfnamefont {K.}~\bibnamefont
  {Wang}}, \bibinfo {author} {\bibfnamefont {A.}~\bibnamefont {Dutt}}, \bibinfo
  {author} {\bibfnamefont {K.~Y.}\ \bibnamefont {Yang}}, \bibinfo {author}
  {\bibfnamefont {C.~C.}\ \bibnamefont {Wojcik}}, \bibinfo {author}
  {\bibfnamefont {J.}~\bibnamefont {Vu{\v{c}}kovi{\'c}}},\ and\ \bibinfo
  {author} {\bibfnamefont {S.}~\bibnamefont {Fan}},\ }\bibfield  {title}
  {\bibinfo {title} {Generating arbitrary topological windings of a
  {non-Hermitian} band},\ }\href
  {https://www.science.org/doi/10.1126/science.abf6568} {\bibfield  {journal}
  {\bibinfo  {journal} {Science}\ }\textbf {\bibinfo {volume} {371}},\ \bibinfo
  {pages} {1240} (\bibinfo {year} {2021})}\BibitemShut {NoStop}%
\bibitem [{\citenamefont {Yuan}\ \emph {et~al.}(2021)\citenamefont {Yuan},
  \citenamefont {Dutt},\ and\ \citenamefont {Fan}}]{Yuan:2021:APL}%
  \BibitemOpen
  \bibfield  {author} {\bibinfo {author} {\bibfnamefont {L.}~\bibnamefont
  {Yuan}}, \bibinfo {author} {\bibfnamefont {A.}~\bibnamefont {Dutt}},\ and\
  \bibinfo {author} {\bibfnamefont {S.}~\bibnamefont {Fan}},\ }\bibfield
  {title} {\bibinfo {title} {Synthetic frequency dimensions in dynamically
  modulated ring resonators},\ }\href
  {https://pubs.aip.org/aip/app/article/6/7/071102/892590/Synthetic-frequency-dimensions-in-dynamically}
  {\bibfield  {journal} {\bibinfo  {journal} {APL Photonics}\ }\textbf
  {\bibinfo {volume} {6}},\ \bibinfo {pages} {071102} (\bibinfo {year}
  {2021})}\BibitemShut {NoStop}%
\bibitem [{\citenamefont {Lustig}\ and\ \citenamefont
  {Segev}(2021)}]{Lustig:2021:Adv}%
  \BibitemOpen
  \bibfield  {author} {\bibinfo {author} {\bibfnamefont {E.}~\bibnamefont
  {Lustig}}\ and\ \bibinfo {author} {\bibfnamefont {M.}~\bibnamefont {Segev}},\
  }\bibfield  {title} {\bibinfo {title} {Topological photonics in synthetic
  dimensions},\ }\href
  {https://opg.optica.org/aop/abstract.cfm?uri=aop-13-2-426} {\bibfield
  {journal} {\bibinfo  {journal} {Advances in Optics and Photonics}\ }\textbf
  {\bibinfo {volume} {13}},\ \bibinfo {pages} {426} (\bibinfo {year}
  {2021})}\BibitemShut {NoStop}%
\bibitem [{\citenamefont {Dutt}\ \emph {et~al.}(2022)\citenamefont {Dutt},
  \citenamefont {Yuan}, \citenamefont {Yang}, \citenamefont {Wang},
  \citenamefont {Buddhiraju}, \citenamefont {Vu{\v{c}}kovi{\'c}},\ and\
  \citenamefont {Fan}}]{Dutt:2022:NatComm}%
  \BibitemOpen
  \bibfield  {author} {\bibinfo {author} {\bibfnamefont {A.}~\bibnamefont
  {Dutt}}, \bibinfo {author} {\bibfnamefont {L.}~\bibnamefont {Yuan}}, \bibinfo
  {author} {\bibfnamefont {K.~Y.}\ \bibnamefont {Yang}}, \bibinfo {author}
  {\bibfnamefont {K.}~\bibnamefont {Wang}}, \bibinfo {author} {\bibfnamefont
  {S.}~\bibnamefont {Buddhiraju}}, \bibinfo {author} {\bibfnamefont
  {J.}~\bibnamefont {Vu{\v{c}}kovi{\'c}}},\ and\ \bibinfo {author}
  {\bibfnamefont {S.}~\bibnamefont {Fan}},\ }\bibfield  {title} {\bibinfo
  {title} {Creating boundaries along a synthetic frequency dimension},\ }\href
  {https://www.nature.com/articles/s41467-022-31140-7} {\bibfield  {journal}
  {\bibinfo  {journal} {Nature Communications}\ }\textbf {\bibinfo {volume}
  {13}},\ \bibinfo {pages} {3377} (\bibinfo {year} {2022})}\BibitemShut
  {NoStop}%
\bibitem [{\citenamefont {Bal{\v{c}}ytis}\ \emph {et~al.}(2022)\citenamefont
  {Bal{\v{c}}ytis}, \citenamefont {Ozawa}, \citenamefont {Ota}, \citenamefont
  {Iwamoto}, \citenamefont {Maeda},\ and\ \citenamefont
  {Baba}}]{Balcytis:2022:SciAdv}%
  \BibitemOpen
  \bibfield  {author} {\bibinfo {author} {\bibfnamefont {A.}~\bibnamefont
  {Bal{\v{c}}ytis}}, \bibinfo {author} {\bibfnamefont {T.}~\bibnamefont
  {Ozawa}}, \bibinfo {author} {\bibfnamefont {Y.}~\bibnamefont {Ota}}, \bibinfo
  {author} {\bibfnamefont {S.}~\bibnamefont {Iwamoto}}, \bibinfo {author}
  {\bibfnamefont {J.}~\bibnamefont {Maeda}},\ and\ \bibinfo {author}
  {\bibfnamefont {T.}~\bibnamefont {Baba}},\ }\bibfield  {title} {\bibinfo
  {title} {Synthetic dimension band structures on a {Si CMOS} photonic
  platform},\ }\href {https://www.science.org/doi/10.1126/sciadv.abk0468}
  {\bibfield  {journal} {\bibinfo  {journal} {Science advances}\ }\textbf
  {\bibinfo {volume} {8}},\ \bibinfo {pages} {eabk0468} (\bibinfo {year}
  {2022})}\BibitemShut {NoStop}%
\bibitem [{\citenamefont {Ehrhardt}\ \emph {et~al.}(2023)\citenamefont
  {Ehrhardt}, \citenamefont {Weidemann}, \citenamefont {Maczewsky},
  \citenamefont {Heinrich},\ and\ \citenamefont {Szameit}}]{Ehrhardt:2023:LPR}%
  \BibitemOpen
  \bibfield  {author} {\bibinfo {author} {\bibfnamefont {M.}~\bibnamefont
  {Ehrhardt}}, \bibinfo {author} {\bibfnamefont {S.}~\bibnamefont {Weidemann}},
  \bibinfo {author} {\bibfnamefont {L.~J.}\ \bibnamefont {Maczewsky}}, \bibinfo
  {author} {\bibfnamefont {M.}~\bibnamefont {Heinrich}},\ and\ \bibinfo
  {author} {\bibfnamefont {A.}~\bibnamefont {Szameit}},\ }\bibfield  {title}
  {\bibinfo {title} {A perspective on synthetic dimensions in photonics},\
  }\href {https://doi.org/https://doi.org/10.1002/lpor.202200518} {\bibfield
  {journal} {\bibinfo  {journal} {Laser \& Photonics Reviews}\ }\textbf
  {\bibinfo {volume} {17}},\ \bibinfo {pages} {2200518} (\bibinfo {year}
  {2023})}\BibitemShut {NoStop}%
\bibitem [{\citenamefont {Li}\ \emph {et~al.}(2023)\citenamefont {Li},
  \citenamefont {Wang}, \citenamefont {Ye}, \citenamefont {Zheng},
  \citenamefont {Wang}, \citenamefont {Liu}, \citenamefont {Dutt},
  \citenamefont {Yuan},\ and\ \citenamefont {Chen}}]{Li:2023:Light}%
  \BibitemOpen
  \bibfield  {author} {\bibinfo {author} {\bibfnamefont {G.}~\bibnamefont
  {Li}}, \bibinfo {author} {\bibfnamefont {L.}~\bibnamefont {Wang}}, \bibinfo
  {author} {\bibfnamefont {R.}~\bibnamefont {Ye}}, \bibinfo {author}
  {\bibfnamefont {Y.}~\bibnamefont {Zheng}}, \bibinfo {author} {\bibfnamefont
  {D.-W.}\ \bibnamefont {Wang}}, \bibinfo {author} {\bibfnamefont {X.-J.}\
  \bibnamefont {Liu}}, \bibinfo {author} {\bibfnamefont {A.}~\bibnamefont
  {Dutt}}, \bibinfo {author} {\bibfnamefont {L.}~\bibnamefont {Yuan}},\ and\
  \bibinfo {author} {\bibfnamefont {X.}~\bibnamefont {Chen}},\ }\bibfield
  {title} {\bibinfo {title} {Direct extraction of topological {Z}ak phase with
  the synthetic dimension},\ }\href
  {https://www.nature.com/articles/s41377-023-01126-1} {\bibfield  {journal}
  {\bibinfo  {journal} {Light: Science \& Applications}\ }\textbf {\bibinfo
  {volume} {12}},\ \bibinfo {pages} {81} (\bibinfo {year} {2023})}\BibitemShut
  {NoStop}%
\bibitem [{\citenamefont {Pellerin}\ \emph {et~al.}(2024)\citenamefont
  {Pellerin}, \citenamefont {Houvenaghel}, \citenamefont {Coish}, \citenamefont
  {Carusotto},\ and\ \citenamefont {St-Jean}}]{Pellerin:PRL2024}%
  \BibitemOpen
  \bibfield  {author} {\bibinfo {author} {\bibfnamefont {F.}~\bibnamefont
  {Pellerin}}, \bibinfo {author} {\bibfnamefont {R.}~\bibnamefont
  {Houvenaghel}}, \bibinfo {author} {\bibfnamefont {W.~A.}\ \bibnamefont
  {Coish}}, \bibinfo {author} {\bibfnamefont {I.}~\bibnamefont {Carusotto}},\
  and\ \bibinfo {author} {\bibfnamefont {P.}~\bibnamefont {St-Jean}},\
  }\bibfield  {title} {\bibinfo {title} {Wave-function tomography of
  topological dimer chains with long-range couplings},\ }\href
  {https://doi.org/10.1103/PhysRevLett.132.183802} {\bibfield  {journal}
  {\bibinfo  {journal} {Phys. Rev. Lett.}\ }\textbf {\bibinfo {volume} {132}},\
  \bibinfo {pages} {183802} (\bibinfo {year} {2024})}\BibitemShut {NoStop}%
\bibitem [{\citenamefont {D'Errico}\ \emph {et~al.}(2020)\citenamefont
  {D'Errico}, \citenamefont {Di~Colandrea}, \citenamefont {Barboza},
  \citenamefont {Dauphin}, \citenamefont {Lewenstein}, \citenamefont
  {Massignan}, \citenamefont {Marrucci},\ and\ \citenamefont
  {Cardano}}]{Derrico:PRR2020}%
  \BibitemOpen
  \bibfield  {author} {\bibinfo {author} {\bibfnamefont {A.}~\bibnamefont
  {D'Errico}}, \bibinfo {author} {\bibfnamefont {F.}~\bibnamefont
  {Di~Colandrea}}, \bibinfo {author} {\bibfnamefont {R.}~\bibnamefont
  {Barboza}}, \bibinfo {author} {\bibfnamefont {A.}~\bibnamefont {Dauphin}},
  \bibinfo {author} {\bibfnamefont {M.}~\bibnamefont {Lewenstein}}, \bibinfo
  {author} {\bibfnamefont {P.}~\bibnamefont {Massignan}}, \bibinfo {author}
  {\bibfnamefont {L.}~\bibnamefont {Marrucci}},\ and\ \bibinfo {author}
  {\bibfnamefont {F.}~\bibnamefont {Cardano}},\ }\bibfield  {title} {\bibinfo
  {title} {Bulk detection of time-dependent topological transitions in quenched
  chiral models},\ }\href {https://doi.org/10.1103/PhysRevResearch.2.023119}
  {\bibfield  {journal} {\bibinfo  {journal} {Phys. Rev. Res.}\ }\textbf
  {\bibinfo {volume} {2}},\ \bibinfo {pages} {023119} (\bibinfo {year}
  {2020})}\BibitemShut {NoStop}%
\bibitem [{\citenamefont {Senanian}\ \emph {et~al.}(2023)\citenamefont
  {Senanian}, \citenamefont {Wright}, \citenamefont {Wade}, \citenamefont
  {Doyle},\ and\ \citenamefont {McMahon}}]{senanian2023programmable}%
  \BibitemOpen
  \bibfield  {author} {\bibinfo {author} {\bibfnamefont {A.}~\bibnamefont
  {Senanian}}, \bibinfo {author} {\bibfnamefont {L.~G.}\ \bibnamefont
  {Wright}}, \bibinfo {author} {\bibfnamefont {P.~F.}\ \bibnamefont {Wade}},
  \bibinfo {author} {\bibfnamefont {H.~K.}\ \bibnamefont {Doyle}},\ and\
  \bibinfo {author} {\bibfnamefont {P.~L.}\ \bibnamefont {McMahon}},\
  }\bibfield  {title} {\bibinfo {title} {Programmable large-scale simulation of
  bosonic transport in optical synthetic frequency lattices},\ }\href
  {https://www.nature.com/articles/s41567-023-02075-7} {\bibfield  {journal}
  {\bibinfo  {journal} {Nature Physics}\ }\textbf {\bibinfo {volume} {19}},\
  \bibinfo {pages} {1333} (\bibinfo {year} {2023})}\BibitemShut {NoStop}%
\bibitem [{\citenamefont {Bergholtz}\ \emph {et~al.}(2021)\citenamefont
  {Bergholtz}, \citenamefont {Budich},\ and\ \citenamefont
  {Kunst}}]{Bergholtz:2021:RMP}%
  \BibitemOpen
  \bibfield  {author} {\bibinfo {author} {\bibfnamefont {E.~J.}\ \bibnamefont
  {Bergholtz}}, \bibinfo {author} {\bibfnamefont {J.~C.}\ \bibnamefont
  {Budich}},\ and\ \bibinfo {author} {\bibfnamefont {F.~K.}\ \bibnamefont
  {Kunst}},\ }\bibfield  {title} {\bibinfo {title} {Exceptional topology of
  {non-Hermitian} systems},\ }\href
  {https://doi.org/10.1103/RevModPhys.93.015005} {\bibfield  {journal}
  {\bibinfo  {journal} {Rev. Mod. Phys.}\ }\textbf {\bibinfo {volume} {93}},\
  \bibinfo {pages} {015005} (\bibinfo {year} {2021})}\BibitemShut {NoStop}%
\bibitem [{\citenamefont {Mizoguchi}\ \emph {et~al.}(2021)\citenamefont
  {Mizoguchi}, \citenamefont {Kuno},\ and\ \citenamefont
  {Hatsugai}}]{Mizoguchi:2021}%
  \BibitemOpen
  \bibfield  {author} {\bibinfo {author} {\bibfnamefont {T.}~\bibnamefont
  {Mizoguchi}}, \bibinfo {author} {\bibfnamefont {Y.}~\bibnamefont {Kuno}},\
  and\ \bibinfo {author} {\bibfnamefont {Y.}~\bibnamefont {Hatsugai}},\
  }\bibfield  {title} {\bibinfo {title} {Detecting bulk topology of quadrupolar
  phase from quench dynamics},\ }\href
  {https://doi.org/10.1103/PhysRevLett.126.016802} {\bibfield  {journal}
  {\bibinfo  {journal} {Phys. Rev. Lett.}\ }\textbf {\bibinfo {volume} {126}},\
  \bibinfo {pages} {016802} (\bibinfo {year} {2021})}\BibitemShut {NoStop}%
\bibitem [{\citenamefont {Wu}\ \emph {et~al.}(2021)\citenamefont {Wu},
  \citenamefont {Guan}, \citenamefont {Fan}, \citenamefont {Chen},\ and\
  \citenamefont {Jia}}]{Wu:2021}%
  \BibitemOpen
  \bibfield  {author} {\bibinfo {author} {\bibfnamefont {C.}~\bibnamefont
  {Wu}}, \bibinfo {author} {\bibfnamefont {X.}~\bibnamefont {Guan}}, \bibinfo
  {author} {\bibfnamefont {J.}~\bibnamefont {Fan}}, \bibinfo {author}
  {\bibfnamefont {G.}~\bibnamefont {Chen}},\ and\ \bibinfo {author}
  {\bibfnamefont {S.}~\bibnamefont {Jia}},\ }\bibfield  {title} {\bibinfo
  {title} {Dynamical characterization of quadrupole topological phases in
  superconducting circuits},\ }\href
  {https://doi.org/10.1103/PhysRevA.104.022601} {\bibfield  {journal} {\bibinfo
   {journal} {Phys. Rev. A}\ }\textbf {\bibinfo {volume} {104}},\ \bibinfo
  {pages} {022601} (\bibinfo {year} {2021})}\BibitemShut {NoStop}%
\bibitem [{\citenamefont {Benalcazar}\ and\ \citenamefont
  {Cerjan}(2022)}]{Benalcazar:2022}%
  \BibitemOpen
  \bibfield  {author} {\bibinfo {author} {\bibfnamefont {W.~A.}\ \bibnamefont
  {Benalcazar}}\ and\ \bibinfo {author} {\bibfnamefont {A.}~\bibnamefont
  {Cerjan}},\ }\bibfield  {title} {\bibinfo {title} {Chiral-symmetric
  higher-order topological phases of matter},\ }\href
  {https://doi.org/10.1103/PhysRevLett.128.127601} {\bibfield  {journal}
  {\bibinfo  {journal} {Phys. Rev. Lett.}\ }\textbf {\bibinfo {volume} {128}},\
  \bibinfo {pages} {127601} (\bibinfo {year} {2022})}\BibitemShut {NoStop}%
\bibitem [{\citenamefont {J{\"u}rgensen}\ \emph {et~al.}(2021)\citenamefont
  {J{\"u}rgensen}, \citenamefont {Mukherjee},\ and\ \citenamefont
  {Rechtsman}}]{Jurgensen:2021}%
  \BibitemOpen
  \bibfield  {author} {\bibinfo {author} {\bibfnamefont {M.}~\bibnamefont
  {J{\"u}rgensen}}, \bibinfo {author} {\bibfnamefont {S.}~\bibnamefont
  {Mukherjee}},\ and\ \bibinfo {author} {\bibfnamefont {M.~C.}\ \bibnamefont
  {Rechtsman}},\ }\bibfield  {title} {\bibinfo {title} {Quantized nonlinear
  {T}houless pumping},\ }\href
  {https://www.nature.com/articles/s41586-021-03688-9} {\bibfield  {journal}
  {\bibinfo  {journal} {Nature}\ }\textbf {\bibinfo {volume} {596}},\ \bibinfo
  {pages} {63} (\bibinfo {year} {2021})}\BibitemShut {NoStop}%
\bibitem [{\citenamefont {Sone}\ \emph {et~al.}(2024)\citenamefont {Sone},
  \citenamefont {Ezawa}, \citenamefont {Ashida}, \citenamefont {Yoshioka},\
  and\ \citenamefont {Sagawa}}]{Sone:2024}%
  \BibitemOpen
  \bibfield  {author} {\bibinfo {author} {\bibfnamefont {K.}~\bibnamefont
  {Sone}}, \bibinfo {author} {\bibfnamefont {M.}~\bibnamefont {Ezawa}},
  \bibinfo {author} {\bibfnamefont {Y.}~\bibnamefont {Ashida}}, \bibinfo
  {author} {\bibfnamefont {N.}~\bibnamefont {Yoshioka}},\ and\ \bibinfo
  {author} {\bibfnamefont {T.}~\bibnamefont {Sagawa}},\ }\bibfield  {title}
  {\bibinfo {title} {Nonlinearity-induced topological phase transition
  characterized by the nonlinear {C}hern number},\ }\href
  {https://www.nature.com/articles/s41567-024-02451-x} {\bibfield  {journal}
  {\bibinfo  {journal} {Nature Physics}\ ,\ \bibinfo {pages} {1}} (\bibinfo
  {year} {2024})}\BibitemShut {NoStop}%
\end{thebibliography}%
\end{document}